\documentstyle[12pt,epsf]{article}
\newcommand{\beq}{\begin{equation}}
\newcommand{\eeq}{\end{equation}}
\newcommand{\beqa}{\begin{eqnarray}}
\newcommand{\eeqa}{\end{eqnarray}}

\newcommand{\Z}{{\bf Z}}

\newcommand{\C}{{\bf C}}
\newcommand{\e}{{\rm e}}
\newcommand{\bra}{\langle}
\newcommand{\ket}{\rangle}
\newcommand{\CP}{\C{\rm P}}
\newcommand{\vsp}{\vspace{0.3cm}}
\newcommand{\Top}{\Omega_{M}}
\newcommand{\TopN}{\Omega_{N}}
\newcommand{\TopMN}{\Omega_{M\times N}}
\newcommand{\tilq}{\tilde{q}}
\newcommand{\cC}{{\cal C}}

\begin{document}
\baselineskip=0.61cm
\normalsize

\thispagestyle{empty}

\begin{flushright}
UT-753  \\
May, 1996 \\
\end{flushright}

\begin{center}
{\Large \bf Gravitational Quantum Cohomology}
\end{center}

\bigskip

\begin{center}

Tohru Eguchi 

\medskip

{\it Department of Physics, University of Tokyo, Tokyo 113, Japan}

\bigskip

Kentaro Hori

\medskip

{\it Institute for Nuclear Study, University of Tokyo, Tokyo 188, Japan}

\bigskip

and

\medskip

Chuan-Sheng Xiong

\medskip

{\it Yukawa Institute for Theoretical Physics, Kyoto University 

\medskip

Kyoto 606, Japan}
\end{center}

\medskip

\begin{center}
{\large \it  Dedicated to the Memory of Claude Itzykson}
\end{center}
\medskip

\begin{abstract}
We discuss how the theory of quantum cohomology may be generalized to 
``gravitational quantum cohomology'' by studying topological $\sigma$-models 
coupled to two-dimensional gravity. We first consider $\sigma$-models defined
on a general Fano manifold $M$ (manifold with a positive first Chern class)
and derive new recursion relations for its two point functions.
We then derive bi-Hamiltonian structures of the theories 
and show that they are completely integrable
at least at the level of genus $0$. We next consider the subspace of the
phase space where only a marginal perturbation (with a parameter 
$t$) is turned on
and construct Lax operators (superpotentials) $L$ whose residue integrals 
reproduce correlation functions. In the case of 
$M=\CP^N$ the Lax operator is given by $L=
Z_1+Z_2+\cdots +Z_N+e^tZ_1^{-1}Z_2^{-1}\cdots Z_N^{-1}$ and agrees with the
potential of the affine Toda theory of the ${\rm A}_N$ type. We also obtain
Lax operators for various Fano manifolds; Grassmannians,
rational surfaces etc.
In these examples the number of variables of the Lax operators is the same 
as the dimension of the original manifold.
Our result shows that Fano manifolds exhibit
a new type of mirror phenomenon where mirror partner is
a non-compact Calabi-Yau manifold of 
the type of an algebraic torus $\C^{*N}$ equipped with a specific 
superpotential.

\end{abstract}

\newpage

\pagenumbering{arabic}
\noindent
{\large \bf 1. Introduction}

\bigskip
The theory of quantum cohomology introduced in \cite{W1,Va}
describes how the quantum effects due to instantons 
modify the classical cohomology of a given manifold $M$.
Relations of the quantum cohomology ring are derived from the study of
correlation functions of a topological sigma model on the sphere
with $M$ being its target space.
The quantum cohomology has been an important arena for the study of
mirror symmetry of Calabi-Yau (CY) manifolds
where the A-twisted superconformal field theory
on one CY is equivalent with the
B-twisted theory on another \cite{Mirror}.
When the target space has a positive first Chern class
(the case of a Fano manifold), 
the underlying sigma model is asymptotically free and has a mass gap,
and one of the two $U(1)_R$ symmetries is anomalously broken.
Accordingly, only A-twisting is possible and the resulting topological 
theory
has an intrinsically broken scale invariance. Nevertheless,
the quantum cohomology of a Fano manifold
can also be described \cite{W1,Va,IV,ST,BeauJin}
by a B-twisted $N=2$ supersymmetric field theory
(topological Landau-Ginzburg (LG) model \cite{Vafa}).
These phenomena, however, will become far more interesting
if the topological sigma model is
coupled with topological gravity \cite{W1},
higher genus contributions are included, and
the gravitational descendants are also incorporated.
Recall that the gravitational descendants (Mumford-Morita classes)
played a prominent role
in the theory of two dimensional gravity \cite{W1,W2,Ko}.
Recently, there have been extensive studies on
topological sigma models coupled with gravity
(or Gromov-Witten invariants)
\cite{KM,I,DI,Dubro,Kon,Curves}
and the rigorous definition is accomplished in \cite{RT,McS},
but the structure of the theory has not yet been well-understood.

In ref.\cite{EY,EHY,Eg} we have considered the simplest Fano manifold
$\CP^1$ and used the standard method of topological field theory 
\cite{W1,W2,DW,H} to analyze the integrable structure of the sigma model
coupled with gravity.
We have constructed a matrix model which reproduces the sum over all 
instantons from Riemann surfaces of arbitrary genus onto $\CP^1$. 
The action of the 
matrix model contains a logarithmic potential which reflects the broken 
scale
invariance of the theory. In ref.\cite{EHY} it was shown that a 
Landau-Ginzburg 
description of the $\CP^1$ model is given by a superpotential of the form 
$\exp x+\exp -x$
and thus the $\CP^1$ model is identified as the $N=2$ sine-Gordon theory.

In this article, at the level of genus 0,
we generalize our construction for a wide class
of Fano varieties including $\CP^N$, Grassmannians, Del Pezzo surfaces and
the products of such spaces.
Based on these constructions we would like to propose a 
theory of ``gravitational quantum cohomology'' which
is a gravitational version of quantum cohomology theory
where the sigma models are coupled with gravity and
the gravitational descendants are incorporated.
We shall show that gravitational quantum cohomology
can be described by a topological LG model
based on a non-compact CY manifold having the same dimension
as the original Fano manifold ($(\C^*)^N$ in the case of $\CP^N$). 
Namely, we can represent the correlation functions of the original model
by residue integrals of a superpotential on the CY.
This indicates the existence of a mirror phenomenon for the
case of Fano varieties as well as in the case of Calabi-Yau manifolds.

Contents of this paper are as follows: in section 2 we derive fundamental 
recursion
relations which relate two-point functions $\langle \sigma_n(O_{\alpha})
O_{\beta}\rangle$ to $\langle \sigma_{n-1}(O_{\alpha})
O_{\gamma}\rangle$
using the machinery of topological field
theory \cite{W1,DW,H}. Here $O_{\alpha}$'s are the primary fields and 
$\sigma_n(O_{\alpha})$ is the $n$-th gravitational descendant of 
$O_{\alpha}$. These relations hold for arbitrary Fano manifolds 
and at general values of the coupling constants (in the large phase space).
In section 3 we derive the
bi-Hamiltonian structure of the topological sigma models
making use of these recursions relations. The existence of the bi-Hamiltonian
structure shows explicitly the complete
integrability of the sigma models on general Fano varieties. 

In section 4 we consider the case of the projective spaces $M=\CP^N$
and discuss
the $\CP^2$ case in detail. We in particular analyze the structure of the 
theory at a point in the phase space where all the 
coupling constants vanish except the one $t$ coupled to the K\"{a}hler 
class (marginal operator). This is the point where the topological 
sigma models lead to quantum cohomology 
relations. In our case the recursion relations among 
two-point functions $\bra\sigma_n(O_{\alpha})O_{\beta}\ket$
become simplified when the coupling constants vanish except $t$ and
the two-point functions can be represented by period integrals 
using a suitable superpotential $L$ (Lax operator). In the case of $\CP^N$ 
$L$ is given by $L=Z_1+Z_2+\cdots+Z_N+e^t Z_1^{-1}Z_2^{-1}\cdots Z_N^{-1}$.
This is the form of the potential of the affine Toda field theory of
${\rm A}_N$ type and is a natural generalization of the sine-Gordon potential 
$L=Z+e^tZ^{-1}$ of the $\CP^1$ 
case. Unlike the case of the quantum cohomology described by one variable 
\cite{Va,IV} we need 
$N$ variables $Z_i, i=1,\cdots,N$ in order to describe the 
gravitational quantum cohomology of $\CP^N$.

Representation of cohomologies by means of period integrals indicate an 
analogue of the mirror phenomenon. 
In the case of $\CP^N$, for instance,
the space $(\C^*)^N$ may be interpreted as 
the mirror manifold of $\CP^N$. The measure of the period integral
$dZ_1dZ_2\cdots dZ_N /(Z_1Z_2\cdots Z_N)$ gives the analogue of the
holomorphic $N$-form. In all the examples we have studied in this paper
the superpotentials of gravitational cohomologies have the same number
of variables as the original manifolds. 
Thus in the case of Fano varieties we have a mirror 
phenomenon where the A model coupled with gravity is equivalent with
the B model on a non-compact CY manifold of the same dimension
(algebraic torus for a toric variety)
equipped with a specific superpotential.

Section 5 is devoted to the study of a product space
where we obtain a sum of superpotentials for each space,
and also some rational surfaces.
In the Appendix A we present a proof
of an important fact concerning the LG description of descendants
and in the Appendix B the Lax operator of some Grassmann manifolds. 

After the completion of this work we have noticed a preprint by Givental
\cite{Givenb} where a mirror phenomenon of a certain class of
Fano manifolds is presented.

\medskip
\noindent Throughout this paper we use the following notations;

$M$:  a K\"ahler manifold with the first Chern class $c_1(M)$,

$\{O_{\alpha}\}$: the base of $H^*(M;\C)$
with $\dim O_{\alpha}=2q_{\alpha}$,

$\displaystyle{\eta_{\alpha \beta}=\int_M O_{\alpha}\wedge O_{\beta}}$:
intersection pairing or the topological metric. 

\noindent
We lower and raise the indices $\alpha,\beta,\ldots$
using the metric $\eta_{\alpha \beta}$ and its inverse $\eta^{\alpha \beta}$.

\vspace{0.7cm}
\renewcommand{\theequation}{2.\arabic{equation}}\setcounter{equation}{0}

\noindent{\large \bf 2. Fundamental Recursion Relation}

\vsp
First, we summarize what is known about topological string
amplitudes
\beq
\bra \sigma_{n_1}(O_{\alpha_1})\cdots\sigma_{n_s}(O_{\alpha_s})\ket_{g,d},
\label{amp}
\eeq
where $g$ is the genus of surfaces and $d$ is the degree of maps which is
defined as the homology class $d=f_*[\Sigma]\in H_2(M)$.

The general relations among correlation functions in the topological string 
theory are given as follows.

\vsp
\noindent $\bullet$ Selection rule (ghost number conservation):
Non-vanishing of (\ref{amp}) requires
\beq
c_1(M)\cdot d+(\dim M-3)(1-g)=\sum_{i=1}^s(n_i+q_{\alpha_i}-1)
\label{selection rule}
\eeq
Here, $c_1(M)\cdot d\in\Z$ is the pairing of $c_1(M)\in H^2(M)$ and $d\in 
H_2(M)$: $c_1(M)\cdot d=\int_{f_*(\Sigma)}c_1(M)=\int_{\Sigma}f^*(c_1(M))$.

\vsp
\noindent $\bullet$ Puncture equation \cite{DW}:
\beq
\bra P \sigma_{n_1}(O_1)\cdots\sigma_{n_s}(O_s)\ket_{g,d}
=\sum_{i=1}^s n_i\bra \sigma_{n_i-1}(O_i)\prod_{j\ne i}\sigma_{n_j}(O_j)
\ket_{g,d}
\label{punc eq}
\eeq

\vsp
\noindent $\bullet$ Dilaton equation \cite{W2}:
\beq
\bra \sigma_1(P) \sigma_{n_1}(O_1)\cdots\sigma_{n_s}(O_s)\ket_{g,d}
=(2g-2+s)\bra \sigma_{n_1}(O_1)\cdots\sigma_{n_s}(O_s)\ket_{g,d}
\label{dila eq}
\eeq

\vsp
\noindent $\bullet$ Equation associated with $\omega\in H^2(M;\C)$
\cite{H}:
\beqa
\lefteqn{\bra \sigma_0(\omega)
\sigma_{n_1}(O_1)\cdots\sigma_{n_s}(O_s)\ket_{g,d}}\label{kahler eq}\\
&&=\omega\cdot d \,\bra \sigma_{n_1}(O_1)\cdots\sigma_{n_s}(O_s)\ket_{g,d}
+\sum_{i=1}^s n_i
\bra \sigma_{n_i-1}(\omega\wedge O_i)\prod_{j\ne i}\sigma_{n_j}(O_j)
\ket_{g,d}\nonumber
\eeqa

\vsp
\noindent $\bullet$ Topological recursion relation (TRR) \cite{W1}:

This is a relation for $g=0$. We denote the sum over the degrees $d$
of $g=0$ amplitudes by
$\bra \cdots \ket=\sum_d\bra \cdots \ket_{0,d}$.
\beq
\bra \sigma_n(O) X Y\ket
=n\bra \sigma_{n-1}(O) O_{\alpha}\ket\eta^{\alpha\beta}
\bra O_{\beta} X Y\ket.
\label{TopRec}
\eeq
Here, $X$ and $Y$ are arbitrary observables.
This holds in the large phase space.

\vsp
In the following we consider only the tree ($g=0$) amplitudes.
In the case of minimal models, TRR is
powerful enough to express correlators of descendants in terms of
correlators of primaries because 0,1, and 2-point functions vanish
at the origin of the phase space.
In the case of topological $\sigma$ models, due to the instanton corrections,
we have non-vanishing 0,1, and 2-point functions to which
TRR cannot be applied.
However, if we use the selection rule (\ref{selection rule})
and the equation (\ref{kahler eq}) for the first Chern class
$c_1(M)$, we can convert an $i$-point function to $i+1$-point function.
For example, let $\Top\in H^{2\dim M}(M)$ be the volume form of
$M$ (whose integral is normalized to $1$), and consider the two point function
$\bra \sigma_n(\Top)O_{\alpha}\ket$ at the origin
of the phase space. Then, we have from (\ref{kahler eq})
\beqa
\bra \sigma_n(\Top)c_1(M)O_{\alpha}\ket&=&
\sum_d\bra\sigma_n(\Top)c_1(M)O_{\alpha}\ket_{0,d}\\
&=&
\sum_d c_1(M)\cdot d\,\,\bra\sigma_n(\Top)O_{\alpha}\ket_{0,d}.
\label{eqa1}
\eeqa
The selection rule (\ref{selection rule}) says that
$\bra\sigma_n(\Top)O_{\alpha}\ket_{0,d}$ is non-vanishing
only when
$c_1(M)\cdot d=n+q_{\alpha}+1$.
Therefore, (\ref{eqa1}) coincides with
$(n+q_{\alpha}+1)\bra\sigma_n(\Top)O_{\alpha}\ket$.
Namely, the two point function
$\bra\sigma_n(\Top)O_{\alpha}\ket$ is proportional to
the three point function
$\bra\sigma_n(\Top)c_1(M)O_{\alpha}\ket$.
If we apply TRR to the latter,
we obtain the following relation
\medskip
\beq
\bra\sigma_n(\Top)O_{\alpha}\ket
=\frac{n}{n+q_{\alpha}+1}\,\bra c_1(M)O_{\alpha}O^{\beta}\ket
\bra\sigma_{n-1}(\Top)O_{\beta}\ket.
\label{prerec1}
\eeq

\vsp
\noindent For a general observable $\sigma_n(O_{\beta})$,
similar relation holds but there is a contribution
coming from the contact term in (\ref{kahler eq}):
\medskip
\beq
\bra\sigma_n(O_{\beta})O_{\alpha}\ket
=\frac{n}{n+\tilq_{\alpha}+\tilq_{\beta}}\Bigl(
\bra c_1(M)O_{\alpha}O^{\gamma}\ket
\bra\sigma_{n-1}(O_{\beta})O_{\gamma}\ket
-\bra\sigma_{n-1}(c_1(M)\wedge O_{\beta})O_{\alpha}\ket\Bigr),
\label{prerec}
\eeq
\medskip
where
$$
\tilq_{\alpha}:=q_{\alpha}+\frac{1-\dim M}{2}.
$$

With more care, these formulae can be generalized
to the following recursion relations
that hold in the large phase space. 
Let us introduce a matrix $M_{\alpha\beta}$ by
\medskip
\beq
M_{\alpha\beta}:=
(q_{\alpha}+q_{\beta}+1-\dim M)\bra O_{\alpha}O_{\beta}\ket
+\int_Mc_1(M)\wedge O_{\alpha} \wedge O_{\beta}.
\label{the M}
\eeq
\medskip
Then, we have for $n,m\geq 1$ the

{\bf Fundamental Recursion Relations}
\medskip
\beqa
\bra\sigma_n(O_{\alpha})O_{\beta}\ket
&=&\frac{n}{n+\tilq_{\alpha}+\tilq_{\beta}}\Bigl(\,
M_{\beta}^{\,\,\,\,\gamma}
\bra\sigma_{n-1}(O_{\alpha})O_{\gamma}\ket
-\bra\sigma_{n-1}(c_1(M)\wedge O_{\alpha})O_{\beta}\ket\,\Bigr),
\label{recrel}\\
\bra\sigma_n(O_{\alpha})\sigma_m(O_{\beta})\ket
&=&
\frac{1}{n+m+\tilq_{\alpha}+\tilq_{\beta}}
\Bigg(\,nm\,M^{\rho\sigma}\bra\sigma_{n-1}(O_{\alpha})O_{\rho}\ket
\bra\sigma_{m-1}(O_{\beta})O_{\sigma}\ket\Biggr.\\
&&\Biggl.-n\,\bra\sigma_{n-1}(c_1(M)\wedge O_{\alpha})\sigma_m(O_{\beta})\ket
-m\,\bra\sigma_n(O_{\alpha})\sigma_{m-1}(c_1(M)\wedge O_{\beta})\ket\,\Biggr).
\nonumber
\eeqa
In particular,
\beq
\bra \sigma_n(\Top)O_{\alpha}\ket
\,=\,\frac{n}{n+q_{\alpha}+1}\,M_{\alpha}^{\,\,\,\,\beta}
\bra\sigma_{n-1}(\Top)O_{\beta}\ket.\label{recrel1}
\eeq

\vsp
\noindent Note that these are powerful relations among correlation functions
which relate two point
functions of descendants to those of primaries.
As compared with the usual case where one inserts $P$,
uses TRR and then integrates
with respect to $x=t_0^P$, the new relations are purely
algebraic and are easier to handle.

\noindent {\bf proof.}\hspace{0.2cm}
The equation (\ref{kahler eq}) for $\omega=c_1(M)$ together with
the selection rule (\ref{selection rule}) yields
\beqa
\lefteqn{\bra c_1(M)
\sigma_{n_1}(O_1)\cdots\sigma_{n_s}(O_s)\ket_{g,d}}\\
&&=(3-\dim M)(1-g) \,\bra \sigma_{n_1}(O_1)\cdots\sigma_{n_s}(O_s)\ket_{g,d}
\nonumber\\
&&+\sum_{i=1}^s\left\{
(n_i+q_i-1)\,\bra \sigma_{n_1}(O_1)\cdots\sigma_{n_s}(O_s)\ket_{g,d}
+n_i \,\bra \sigma_{n_i-1}(c_1(M)\wedge O_i)\prod_{j\ne i}\sigma_{n_j}(O_j)
\ket_{g,d}\right\},\nonumber
\eeqa
which is equivalent at $g=0$ to the following equation for the free energy
\beqa
\bra c_1(M) \ket&=&
(3-\dim M)F_0
+\sum_{m,\sigma}\left\{
(m+q_{\sigma}-1)t_m^{\sigma}\frac{\partial}{\partial t_m^{\sigma}}
+m t_m^{\sigma}
c_1(M)_{\,\,\sigma}^{\rho}\frac{\partial}{\partial t_{m-1}^{\rho}}
\right\}F_0\nonumber\\
&&+\,\frac{1}{2}\,t_0^{\sigma}t_0^{\rho}
\int_Mc_1(M)\wedge O_{\sigma}\wedge O_{\rho}.
\eeqa
The last term comes from the degree zero contribution to three
point functions to which the formula (\ref{kahler eq}) cannot be applied.
Let us introduce the perturbed first Chern class
\beq
\cC_1(M):=c_1(M)
-\sum_{m,\sigma}(m+q_{\sigma}-1)t_m^{\sigma}\sigma_m(O_{\sigma})
-\sum_{m,\sigma}mt_m^{\sigma}\sigma_{m-1}(c_1(M)\wedge O_{\sigma}).
\eeq
Then, the above equation is neatly expressed as
\beq
\bra\cC_1(M)\ket
=(3-\dim M)F_0+\,\frac{1}{2}\,t_0^{\sigma}t_0^{\rho}
\int_Mc_1(M)\wedge O_{\sigma}\wedge O_{\rho}.
\label{neat}
\eeq
Taking the derivative of this equation
with respect to $t_n^{\alpha}$ and $t_0^{\beta}$,
we get
\beqa
\bra \cC_1(M) \sigma_n(O_{\alpha})O_{\beta}\ket\!\!\!
&-&\!\!\!(n+q_{\alpha}+q_{\beta}-2)\bra\sigma_n(O_{\alpha})O_{\beta}\ket
-n\,\bra\sigma_{n-1}(c_1(M)\wedge O_{\alpha})O_{\beta}\ket
\nonumber\\
=&&\hspace{-0.8cm}
(3-\dim M)\bra\sigma_n(O_{\alpha})O_{\beta}\ket
+\delta_{n,0}\int_M c_1(M)\wedge O_{\alpha}\wedge O_{\beta}.
\eeqa
For $n=0$, this gives
\beq
\bra \cC_1(M) O_{\alpha}O_{\beta}\ket
=M_{\alpha\beta}.
\label{Mis3}
\eeq
For $n>0$, we have
\beq
\bra\cC_1(M)\sigma_n(O_{\alpha})O_{\beta}\ket
=
(n+q_{\alpha}+q_{\beta}+1-\dim M)
\bra \sigma_n(O_{\alpha})O_{\beta}\ket
+n\,\bra\sigma_{n-1}(c_1(M)\wedge O_{\alpha})O_{\beta}\ket,
\eeq
while it follows from the topological recursion relation that
\beq
\bra\cC_1(M)\sigma_n(O_{\alpha})O_{\beta}\ket
=
n\,\bra\sigma_{n-1}(O_{\alpha})O_{\gamma}\ket
\bra O^{\gamma}\cC_1(M) O_{\beta}\ket.
\eeq
Combing the above three equations, we obtain
the recursion formula (\ref{recrel}).
Proof of the other relation is similar. 

\vsp
\noindent{\it Remark.}
With only the primary marginal perturbation,
we have $\cC_1(M)=c_1(M)$, and therefore
\beq
M_{\alpha\beta}=\bra c_1(M)O_{\alpha}O_{\beta}\ket
\qquad
\mbox{for $t_n^{\gamma}=0$ unless
$n=0$ and $q_{\gamma}=1$.}
\eeq
Then, the relations (\ref{recrel}) and (\ref{recrel1})
coincides with the ones (\ref{prerec}) and (\ref{prerec1})
obtained previously.

\newpage
\vspace{0.7cm}
\renewcommand{\theequation}{3.\arabic{equation}}\setcounter{equation}{0}

\noindent{\large \bf 3. Bi-Hamiltonian Structure}

\bigskip
Every topological string theory can be considered,
at least at the tree level, as an integrable system
where the two point functions constitute the densities for
commuting Hamiltonians.
Here we study the structure of the integrable system
using the fundamental recursion relation (\ref{recrel})
among the Hamiltonian densities.
In particular, we determine the bi-Hamiltonian structure of the topological
$\sigma$ models.

\bigskip
\noindent
{\sc 3.1 The Bi-Hamiltonian Structure}

\vsp
We take $x=t_0^P$ as the basic (or "spatial") coordinate and regard other
coupling constants as infinitely many "time" coordinates.
The order parameters of the theory are defined by
\beq
u_{\alpha}:=\bra PO_{\alpha}\ket.
\eeq
Other two point functions are regarded as functions of these order parameters
("constitutive relations" \cite{DW}).
The particular two point functions with a puncture insertion
\beq
R_{n,\alpha}:=\bra\sigma_n(O_{\alpha})P\ket
\eeq
generates flows in the phase space
in the following sense:

\bigskip
\noindent {\sc First Hamiltonian Structure}

\medskip
With respect to the first Poisson bracket
\beq
\{u_{\alpha}(x),u_{\beta}(y)\}_1=\eta_{\alpha\beta}\partial_x\delta(x-y)
\label{poisson1}
\eeq
$R_{n+1,\alpha}$ acts as the Hamiltonian for the evolution in the 
variable $t_n^{\alpha}$
\beq
\frac{\partial u_{\beta}}{\partial t_n^{\alpha}}
=\frac{1}{n+1}\Bigl\{\,u_{\beta},\int R_{n+1,\alpha} {\rm d} x\,\Bigr\}_1
\label{flow1}
\eeq

\bigskip
\noindent {\sc Second Hamiltonian Structure}

\medskip
With respect to the second Poisson bracket
\beq
\{u_{\alpha}(x),u_{\beta}(y)\}_2
=
\Bigl(\,
M_{\alpha\beta}\partial_x
+\tilq_{\beta}\bra O_{\alpha}O_{\beta}\ket^{\prime}\,\Bigr)\delta(x-y).
\label{poisson2}
\eeq
$R_{n,\alpha}$ generates the evolution in the parameter $t_n^{\alpha}$
\beq
\frac{\partial u_{\beta}}{\partial t_n^{\Top}}
=\frac{1}{n+\frac{\dim M+1}{2}}
\Bigl\{\,u_{\beta},\int R_{n,\Top}{\rm d}x\,\Bigr\}_2\,,
\label{flow2}
\eeq
\beq
\frac{\partial u_{\beta}}{\partial t_n^{\alpha}}
+\frac{n}{n+\tilq_{\alpha}}c_1(M)_{\alpha}^{\,\,\gamma}
\frac{\partial u_{\beta}}{\partial t_{n-1}^{\gamma}}
=
\frac{1}{n+\tilq_{\alpha}}
\Bigl\{\,u_{\beta},\int R_{n,\alpha}{\rm d}x\,\Bigr\}_2.
\label{flow3}
\eeq

\bigskip
Below, we present a derivation and consistency check of these Hamiltonian
structures.

The flow equation (\ref{flow1}) is a consequence of a lemma

\noindent {\bf Lemma 1.}
\beq
\eta_{\beta\gamma}\frac{\partial}{\partial u_{\gamma}}R_{n,\alpha}
=n\,\bra\sigma_{n-1}(O_{\alpha})O_{\beta}\ket
\label{lem1}
\eeq
{\bf proof.} From the constitutive relation it follows that
\beq
\bra\sigma_n(O_{\alpha})PO_{\rho}\ket
=\frac{\partial u_{\gamma}}{\partial t_0^{\rho}}
\frac{\partial}{\partial u_{\gamma}}R_{n,\alpha}
=\bra O_{\rho}PO_{\gamma}\ket
\frac{\partial}{\partial u_{\gamma}}R_{n,\alpha},
\eeq
while TRR gives
\beq
\bra\sigma_n(O_{\alpha})PO_{\rho}\ket
=n\,\bra\sigma_{n-1}(O_{\alpha})O_{\beta}\ket
\bra O^{\beta}PO_{\rho}\ket.
\eeq
Connecting the above two, and multiplying the inverse matrix of
$\bra O^{\beta}PO_{\rho}\ket$, we see that the equation (\ref{lem1}) holds.

\vsp
Let us introduce a convenient notation
\beq
P_{\alpha\beta}:=\bra PO_{\alpha}O_{\beta}\ket.
\eeq
Then, we have

\noindent{\bf Lemma 2.}
{\it The matrices $M$ and $P$ commute}:
\beq
P_{\alpha}^{\,\,\gamma}M_{\gamma \beta}=
M_{\alpha}^{\,\,\gamma}P_{\gamma \beta}.
\label{commu}
\eeq
{\bf proof.} Since $M_{\alpha\beta}$ can be written as a three point
function $\bra\cC_1(M)O_{\alpha}O_{\beta}\ket$ (see (\ref{Mis3})),
the lemma follows from the associativity relation
$\bra XYO^{\gamma}\ket\bra O_{\gamma}ZW\ket=
\bra XZO^{\gamma}\ket\bra O_{\gamma}YW\ket$
which holds for arbitrary observables $X,Y,Z,W$.

\vsp
\noindent
Now, we present a derivation of the second Poisson bracket (\ref{poisson2})
and the flow equation (\ref{flow2}). From the fundamental recursion 
relation (\ref{recrel1}),
we have $(n+q_{\gamma})\bra\sigma_{n-1}(\Top)O_{\gamma}\ket
=(n-1)M_{\gamma}^{\,\,\beta}\bra\sigma_{n-2}(\Top)O_{\beta}\ket$.
Multiplying $P_{\alpha}^{\,\,\gamma}$ we get
\beqa
\sum_{\gamma}P_{\alpha}^{\,\,\gamma}(n+q_{\gamma})
\bra\sigma_{n-1}(\Top)O_{\gamma}\ket
&=&
(n-1)P_{\alpha}^{\,\,\gamma}
M_{\gamma}^{\,\,\beta}\bra\sigma_{n-2}(\Top)O_{\beta}\ket\\
&=&(n-1)M_{\alpha}^{\,\,\gamma}
P_{\gamma}^{\,\,\beta}\bra\sigma_{n-2}(\Top)O_{\beta}\ket\\
&=&
M_{\alpha}^{\,\,\gamma}\bra\sigma_{n-1}(\Top)O_{\gamma}P\ket,
\eeqa
where we used the commutativity (\ref{commu}) in the second step
and the TRR in the last step. 
Using the TRR in the left hand side,
we find that
\beq
\bra\sigma_n(\Top)O_{\alpha}P\ket
=\sum_{\gamma}
\Bigl(M_{\alpha}^{\,\,\gamma}\partial_x
-P_{\alpha}^{\,\,\gamma}q_{\gamma}
\Bigr)
\bra\sigma_{n-1}(\Top)O_{\gamma}\ket
\eeq
Multiplying by a factor $n+\xi$
($\xi$ is a constant to be determined later), we obtain
\beqa
(n+\xi)\bra\sigma_n(\Top)O_{\alpha}P\ket&=&
\sum_{\gamma}\Bigl(M_{\alpha}^{\,\,\gamma}\partial_x
-P_{\alpha}^{\,\,\gamma}q_{\gamma}
\Bigr)
n\,\bra\sigma_{n-1}(\Top)O_{\gamma}\ket
+\xi \,n P_{\alpha}^{\,\,\gamma}\bra\sigma_{n-1}(\Top)O_{\gamma}\ket
\nonumber\\
&=&\sum_{\gamma}\Bigl(M_{\alpha}^{\,\,\gamma}\partial_x
+P_{\alpha}^{\,\,\gamma}(\xi-q_{\gamma})
\Bigr)\eta_{\gamma\beta}
\frac{\partial}{\partial u_{\beta}}R_{n,\Top}
\nonumber\\
&=&
\sum_{\beta}
\Bigl(M_{\alpha\beta}\partial_x
+P_{\alpha\beta}(\xi-\dim M+q_{\beta})
\Bigr)
\frac{\partial}{\partial u_{\beta}}R_{n,\Top},
\eeqa
where we used the TRR in the first step
and the Lemma 1 in the second step.
In the last step, we have used the fact that
$\eta_{\gamma\beta}\ne 0$ only if $q_{\gamma}+q_{\beta}=\dim M$.
If we can define a Poisson bracket $\{\,\,,\,\,\}$ by
\beq
\{u_{\alpha}(x),u_{\beta}(y)\}
=\Bigl(M_{\alpha\beta}\partial_x
+P_{\alpha\beta}(\xi-\dim M+q_{\beta})\Bigr)\delta(x-y),
\label{prepoisson}
\eeq
then the above equation will be expressed as
\beq
\frac{\partial}{\partial t_n^{\Top}}u_{\alpha}=
\frac{1}{n+\xi}\Bigl\{\,u_{\alpha},\int R_{n,\Top}{\rm d}x\,\Bigr\}.
\label{preflow}
\eeq
For (\ref{prepoisson}) to be a Poisson bracket,
it must be anti-symmetric and satisfy the Jacobi identity.
Anti-symmetry requires
\beq
P_{\alpha\beta}(\xi-\dim M+q_{\beta})
+P_{\beta\alpha}(\xi-\dim M+q_{\alpha})
=M_{\alpha\beta}^{\prime}
=(q_{\alpha}+q_{\beta}+1-\dim M)P_{\alpha\beta},
\eeq
or
\beq
\xi=\frac{\dim M+1}{2}.
\eeq
If we plug this into (\ref{prepoisson}) and (\ref{preflow}),
we get (\ref{poisson2}) and (\ref{flow2}).

Finally, let us check the Jacobi identity.
For a test function $a^{\alpha}(x)$, we put
\beq
u[a]:=\int u_{\alpha}a^{\alpha}{\rm d}x.
\eeq
Let $a,b,$ and $c$ be test functions.
\beqa
\{u[a],\{u[b],u[c]\}\}&=&
\Bigl\{u[a],\int{\rm d}x \,b^{\beta} \Bigl(M_{\beta\gamma}c^{\gamma\prime}
+\bra O_{\beta}O_{\gamma}\ket^{\prime}\tilq_{\gamma}c^{\gamma}\Bigr)\Bigr\}
\nonumber\\
&=&\Bigl\{u[a],\int{\rm d}x\Bigl(M_{\beta\gamma}b^{\beta}c^{\gamma\prime}
-\bra O_{\beta}O_{\gamma}\ket\tilq_{\gamma}(b^{\beta}c^{\gamma})^{\prime}
\Bigr)\Bigr\}\nonumber\\
&=&
\int{\rm d}x\,a^{\alpha}
\Bigl(M_{\alpha\rho}\partial_x+\bra O_{\alpha}O_{\rho}\ket^{\prime}\tilq_{\rho}\Bigr)
\left(
\frac{\partial M_{\beta\gamma}}{\partial u_{\rho}}b^{\beta}c^{\gamma\prime}
-\frac{\partial \bra O_{\beta}O_{\gamma}\ket}{\partial u_{\rho}}
\tilq_{\gamma}(b^{\beta}c^{\gamma})^{\prime}\right)
\nonumber\\
=\int{\rm d}x\hspace{0.2cm}&&\hspace{-1.2cm}
\Bigl(-a^{\alpha\prime}M_{\alpha\rho}-a^{\alpha}M_{\alpha\rho}^{\,\,\prime}
+a^{\alpha}\bra O_{\alpha}O_{\rho}\ket^{\prime}\tilq_{\rho}\Bigr)
\left(
\frac{\partial M_{\beta\gamma}}{\partial u_{\rho}}b^{\beta}c^{\gamma\prime}
-\frac{\partial \bra O_{\beta}O_{\gamma}\ket}{\partial u_{\rho}}
\tilq_{\gamma}(b^{\beta}c^{\gamma})^{\prime}\right).
\nonumber
\eeqa
Here we note that
$M_{\alpha\beta}=(\tilq_{\alpha}+\tilq_{\beta})\bra O_{\alpha}O_{\beta}\ket
+{\rm const}$. So,
\beqa
&&-a^{\alpha}M_{\alpha\rho}^{\,\,\prime}
+a^{\alpha}\bra O_{\alpha}O_{\rho}\ket^{\prime}\tilq_{\rho}
=-a^{\alpha}\tilq_{\alpha}\bra O_{\alpha}O_{\rho}\ket^{\prime},\\
&&
\frac{\partial M_{\beta\gamma}}{\partial u_{\rho}}b^{\beta}c^{\gamma\prime}
-\frac{\partial \bra O_{\beta}O_{\gamma}\ket}{\partial u_{\rho}}
\tilq_{\gamma}(b^{\beta}c^{\gamma})^{\prime}=
\frac{\partial \bra O_{\beta}O_{\gamma}\ket}{\partial u_{\rho}}
(\tilq_{\beta}b^{\beta}c^{\gamma\prime}-
\tilq_{\gamma}b^{\beta\prime}c^{\gamma}).
\eeqa
Let us put
\beq
M_{\alpha\beta\gamma}
:=M_{\alpha\rho}
\frac{\partial \bra O_{\beta}O_{\gamma}\ket}{\partial u_{\rho}},
\eeq
and note that
\beq
\bra O_{\alpha}O_{\beta}O_{\gamma}\ket=
\bra O_{\alpha}O_{\rho}\ket^{\prime}
\frac{\partial \bra O_{\beta}O_{\gamma}\ket}{\partial u_{\rho}}.
\eeq
Then, we have
\beqa
\lefteqn{\{u[a],\{u[b],u[c]\}\}}\label{uabc}\\
&=&-\int{\rm d}x
\Big(\,
M_{\alpha\beta\gamma}
(\tilq_{\beta}a^{\alpha\prime}b^{\beta}c^{\gamma\prime}
-\tilq_{\gamma}a^{\alpha\prime}b^{\beta\prime}c^{\gamma})
+\bra O_{\alpha}O_{\beta}O_{\gamma}\ket
(\tilq_{\alpha}\tilq_{\beta}a^{\alpha}b^{\beta}c^{\gamma\prime}
-\tilq_{\alpha}\tilq_{\gamma}a^{\alpha}b^{\beta\prime}c^{\gamma})
\Bigr).\nonumber
\eeqa
{\bf Lemma 3.}
{\it $M_{\alpha\beta\gamma}$ is symmetric}.

\noindent{\bf proof.}
In the small phase space,
$\frac{\partial \bra O_{\beta}O_{\gamma}\ket}{\partial u_{\rho}}
=\bra O^{\rho}O_{\beta}O_{\gamma}\ket$, and hence
\beqa
M_{\alpha\rho}
\frac{\partial \bra O_{\beta}O_{\gamma}\ket}{\partial u_{\rho}}
&=&
\bra O_{\alpha} \cC_1(M) O_{\rho}\ket
\bra O^{\rho}O_{\beta}O_{\gamma}\ket\nonumber\\
&=&
\bra O_{\beta} \cC_1(M) O_{\rho}\ket
\bra O^{\rho}O_{\alpha}O_{\gamma}\ket
=
M_{\beta\rho}
\frac{\partial \bra O_{\alpha}O_{\gamma}\ket}{\partial u_{\rho}}.
\eeqa
Namely, $M_{\alpha\beta\gamma}=M_{\beta\alpha\gamma}$.
As this is a relation of two point functions,
it holds also in the large phase space.

By looking at the expression (\ref{uabc}) we note the symmetry of
$M_{\alpha\beta\gamma}$ and $\bra O_{\alpha}O_{\beta}O_{\gamma}\ket$,
and find that the Jacobi identity
\beq
\{u[a],\{u[b],u[c]\}\}+\{u[b],\{u[c],u[a]\}\}+\{u[c],\{u[a],u[b]\}\}=0
\eeq
holds.

\bigskip
\noindent
{\sc 3.2 Examples}

\bigskip
\noindent
{\it The $\CP^1$ Model}

\medskip
The $\CP^1$ model has two primaries $P$ and $Q$ corresponding to
the identity and the K\"ahler form (normalized to have a unit volume).
The metric is given by $\eta_{PQ}=1$ and $\eta_{PP}=\eta_{QQ}=0$.
Order parameters are denoted as
\beq
u=\bra PP\ket,\qquad v=\bra PQ\ket.
\eeq
The remaining primary two point function $\bra QQ\ket$
is expressed in terms of an order parameter by
the constitutive relation \cite{DW}
\beq
\bra QQ\ket =\e^u.
\eeq
The fundamental matrix
$(M_{\alpha}^{\,\,\beta})=(M_{\alpha\gamma})(\eta^{\gamma\beta})$
is then expressed as
\beq
\left(
\begin{array}{cc}
v& 2\\
2\e^u&v
\end{array}
\right)
\label{MCP1}
\eeq
(with respect to the order $P$,$Q$).
Using this matrix, we can recursively construct all the two point functions.
Applying (\ref{recrel1}), we have for example
\beqa
v_n:=\left(
\begin{array}{c}
\!\!\bra \sigma_n(Q)P\ket\!\!\\
\!\!\bra \sigma_n(Q)Q\ket\!\!
\end{array}
\right):&&\!\!\!\!
v_0
=
\left(
\begin{array}{c}
v\\
\e^u
\end{array}
\right),
v_1
=
\left(
\begin{array}{c}
\frac{v^2}{2}+\e^u\\
v\e^u
\end{array}
\right),
v_2
=
\left(
\begin{array}{c}
\frac{v^3}{3}+2v\e^u\\
v^2\e^u+\e^{2u}
\end{array}
\right),
\label{twoptCP1}\\
&&\!\!\!\!
v_3
=
\left(
\begin{array}{c}
\frac{v^4}{4}+3v^2\e^u+\frac{3}{2}\e^{2u}\\
v^3\e^u+3v\e^{2u}
\end{array}
\right),
v_4
=
\left(
\begin{array}{c}
\frac{v^5}{5}+4v^3\e^u+6v\e^{2u}\\
v^4\e^u+6v^2\e^{2u}+2\e^{3u}
\end{array}
\right),\ldots
\nonumber
\eeqa
The second Poisson bracket is expressed as
\beq
\left(
\begin{array}{cc}
\{u(x),u(y)\}&\{u(x),v(y)\}\\
\{v(x),u(y)\}&\{v(x),v(y)\}
\end{array}
\right)
=
\left(
\begin{array}{cc}
2\partial_x&v\partial_x+v^{\prime}\\
v\partial_x&2\e^u\partial_x+\e^uu^{\prime}
\end{array}
\right)\delta(x-y)\,.
\eeq

\bigskip
\noindent
{\it The $\CP^2$ model}

\medskip
The $\CP^2$ model has three primaries $P,Q,R$ corresponding to
$1,\omega,\omega^2\in H^*(\CP^2)$, respectively
where $\omega\in H^*(\CP^2)$ is the K\"ahler class such that $\omega^2$
has a unit volume.
The metric is given by $\eta_{PR}=\eta_{QQ}=1$, the others $=0$.
Order parameters are denoted as
\beq
u=\bra PP\ket,\quad v=\bra PQ\ket,\quad w=\bra PR\ket.
\eeq
Other two point functions of primaries are essentially the derivatives
of a function $f(u,v)$
\beq
\bra QQ\ket=w+f_{vv},\quad \bra QR\ket=f_{uv},\quad
\bra RR\ket=f_{uu}.
\eeq
Here, $f$ is the instanton contribution to
the free energy:
\beq
f(u,v)=\sum_{d=1}^{\infty}N_d\,\frac{u^{3d-1}}{(3d-1)!}\,\e^{dv},
\eeq
which is determined \cite{KM,I,Dubro}
by the associativity equation \cite{W1,W2}
(or ``WDVV equation'')
\beq
f_{uuu}=(f_{uvv})^2-f_{uuv}f_{vvv},
\eeq
together with the initial value $N_1=1$. For example,
\beq
N_1=1,\quad N_2=1,\quad N_3=12,\quad N_4=620,\quad N_5=87304,\,.\,.\,.
\eeq
The fundamental $M$ matrix is then expressed as
\beq
\left(
\begin{array}{ccc}
w&3&-u\\
2f_{uv}&w+f_{vv}&3\\
3f_{uu}&2f_{uv}&w
\end{array}
\right).
\label{MCP2}
\eeq
The 2nd Poisson bracket is given by
\begin{eqnarray}
&&\left(
\begin{array}{ccc}
\!\{u(x),u(y)\}\!&\!\{u(x),v(y)\}\!&\!\{u(x),w(y)\}\!\\
\!\{v(x),u(y)\}\!&\!\{v(x),v(y)\}\!&\!\{v(x),w(y)\}\!\\
\!\{w(x),u(y)\}\!&\!\{w(x),v(y)\}\!&\!\{w(x),w(y)\}\!
\end{array}\right)  \\
&&\hspace{3mm}=
\left(
\begin{array}{ccc}
-u\partial_x-\frac{1}{2}u^{\prime}&3\partial_x+\frac{1}{2}v^{\prime}
&w\partial_x+\frac{3}{2}w^{\prime}\\
3\partial_x-\frac{1}{2}v^{\prime}&
\!(w+f_{vv})\partial_x+\frac{1}{2}(w+f_{vv})^{\prime}\!
&2f_{uv}\partial_x+\frac{3}{2}f_{uv}^{\,\,\prime}\\
w\partial_x-\frac{1}{2}w^{\prime}&
2f_{uv}\partial_x+\frac{1}{2}f_{uv}^{\,\,\prime}&
3f_{uu}\partial_x+\frac{3}{2}f_{uu}^{\,\,\prime}
\end{array}
\right)\delta(x-y).
\end{eqnarray}

\bigskip
\noindent
{\it Minimal Models}

\medskip
The $k^{\rm th}$ minimal model
can formally be considered as
a topological string theory with a fictitious ``target space'' $M_k$
of dimension $k/(k+2)$
and cohomology classes $O_{\alpha}$ ($\alpha=0,1,\ldots,k$)
of dimensions $\alpha/(k+2)$.
The first Chern class of $M_k$ is assumed to vanish $c_1(M_k)=0$.
Namely, the correlation functions obey the basic equations
(the selection rule, topological recursion relation,
puncture and dilaton equation)
as if the model had such a target space.
The fundamental recursion relation (\ref{recrel})
and the formulae (\ref{poisson1})-(\ref{flow2})
for the bi-Hamiltonian structure apply also to this case,
since the derivation
only needs the selection rule and TRR
in a theory with scale invariance.

For example, let us consider the case $k=0$ of pure topological gravity.
There is only a single primary $P$ and the order parameter is denoted as
$u=\bra PP\ket$.
The fundamental matrix is given by
$M_{0}^{\,\,0}=M_{0 \,0}=u$ and hence the recursion relation (\ref{recrel})
reads as
\beq
\bra \sigma_n(P) P\ket=\frac{n}{n+1}u \,\bra\sigma_{n-1}(P) P\ket.
\eeq
This yields
\beq
\bra \sigma_n(P) P\ket=\frac{1}{n+1}u^{n+1},
\eeq
which is the well-known constitutive relation \cite{DW}.
The second Poisson bracket (\ref{poisson2}) is expressed as
\beq
\{u(x),u(y)\}_2
=\left(\,u(x)\partial_x+\frac{1}{2}u^{\prime}(x)\,\right)
\delta(x-y).
\eeq

\medskip
\noindent
We see that this coincides with
the second Poisson bracket of the full KdV hierarchy
\beq
\{u(x),u(y)\}
=\left(\,-\frac{1}{4}\partial_x^3+
u(x)\partial_x+\frac{1}{2}u^{\prime}(x)\,\right)
\delta(x-y),
\eeq
in the dispersionless limit $\partial_x^3\to 0$.

\medskip
\noindent
{\it Remark}.
The fundamental recursion relation
(\ref{recrel}) in the minimal model
has never been noted in the previous studies of
generalized KdV hierarchy.

\bigskip
\noindent
{\sc 3.3 Virasoro Algebra}

\medskip
So far, we have been studying the integrable hierarchy by taking
$t_0^{P}$ as the basic spatial coordinate.
At least formally, however, we may regard the variable $t_0^{\alpha}$
of an arbitrary primary field $O_{\alpha}$ as the basic coordinate.
(``Democracy'' among the primary fields in minimal models
is discussed in \cite{AoKo}.)

Let us take as the basic spatial coordinate $x$
the coupling constant $t_0^{\Top}$
for the volume class $\Top$.
The class $\Top$ has the property
$c_1(M)\wedge \Top=0$ and
we expect some restoration of scale invariance
as is pointed out in \cite{H} in a different context.
The order parameters are defied by
$U_{\alpha}=\bra \Top O_{\alpha}\ket$.
The formulae (\ref{poisson1})-(\ref{flow2})
for the bi-Hamiltonian structure still hold,
provided we make the replacement $u_{\alpha}\to U_{\alpha}$
and $R_{n,\alpha}=\bra \sigma_n(O_{\alpha})P\ket\to
\bra \sigma_n(O_{\alpha})\Top\ket$.
Here, we note that the second Poisson bracket of the two point function
\beq
T:=\frac{2}{\dim M+1}U_{\Top}=\frac{2}{\dim M+1}\bra \Top\Top\ket
\eeq
gives nothing but the commutation relation of the Virasoro algebra
\beq
\{T(x),T(y)\}_2=(2T(x)\partial_x+T^{\prime}(x))\delta(x-y).
\eeq
Other order parameters $U_{\alpha}$ become the "primary fields"
\beq
\{U_{\alpha}(x),T(y)\}_2=
\left(\,\frac{2(q_{\alpha}+1)}{\dim M+1}U_{\alpha}(x)\partial_x
+U_{\alpha}^{\prime}(x)\,\right)\delta(x-y),
\eeq
of dimensions
$2(q_{\alpha}+1)/(\dim M+1)$.

\newpage
\vspace{0.7cm}
\renewcommand{\theequation}{4.\arabic{equation}}\setcounter{equation}{0}

\noindent{\large \bf 4. Lax Operator and Landau-Ginzburg Formulation}

\bigskip
The fundamental recursion relation (\ref{recrel})
completely determines the structure of the integrable hierarchy
of the tree-level topological string theory.
The integrable system for the full theory
including higher genera will be
a quantization or central extension of the tree level theory.
Conventional way to achieve this is to formulate
a Lax pair representation of the tree level system,
and then to quantize it or discretize it
by finding matrix integral representation.
In this paper, we make a modest step toward the formulation
of Lax pair representation.
In particular, in this section
we determine the Lax operator for the $\CP^N$ model at the tree level
in the relevant and marginal (K\"ahler) perturbations.
We shall also see that we can develop a Landau-Ginzburg description
of the system by regarding the Lax operator as its superpotential.
In the next section,
we treat the case of some other target spaces.

\bigskip
\noindent
{\sc 4.1 A Review Of The $\CP^1$ Model}

\medskip
First, we briefly review the Lax pair representation for the tree level
theory of the $\CP^1$ model \cite{EY}.
As in \S 3,
the order parameters are denoted as
$u=\bra PP\ket$ and $v=\bra PQ\ket$.
The Lax operator is given by
\beq
L=p+v+e^up^{-1}.
\eeq
Here, $p$ is the momentum variable in the dispersionless
Toda lattice hierarchy in which the Poisson bracket is defined by
\beq
\{A,B\}=p\frac{\partial A}{\partial p}\frac{\partial B}{\partial t_0^P}
-p\frac{\partial B}{\partial p}\frac{\partial A}{\partial t_0^P}.
\eeq
The flow equation (\ref{flow1}) or (\ref{flow2}),(\ref{flow3})
can be written in the Lax form
\beq
\frac{\partial L}{\partial t_n^{\alpha}}=
\Bigl\{\Bigl(G_{n,\alpha}(L)\Bigr)_+,L\Bigr\},
\label{laxform}
\eeq
where $(\cdots)_+$ means to take the non-negative powers of $p$,
and
$G_{n,P}$ and $G_{n,Q}$ are given by
\beqa
G_{n,Q}&=&\frac{1}{n+1}L^{n+1},
\label{GnQCP1}\\[0.2cm]
G_{n,P}&=&2L^n(\log L-c_n).
\label{GnPCP1}
\eeqa
In the second expression, $c_n$ is given by
\beq
c_n=1+\frac{1}{2}+\cdots +\frac{1}{n},
\label{cn}
\eeq
and the logarithm of $L$ is defined as a Laurent series in $p$
by taking the average
\begin{eqnarray}
&&\log L=\frac{1}{2}\left(\,
\log\Bigl\{ p(1+vp^{-1}+\e^u p^{-2})\Bigr\}
+\log\Bigl\{ \e^u p^{-1}(\e^{-u}p^2+v\e^{-u}p+1)\Bigr\}\,\right) \\
&&\hspace{5mm}=\frac{1}{2}\left(\,
u+\log (1+vp^{-1}+\e^u p^{-2})
+\log (1+v\e^{-u}p+\e^{-u}p^2)\right).
\end{eqnarray} From (\ref{laxform}) we have
\beqa
\frac{\partial u}{\partial t_n^{\alpha}}&=&
\frac{\partial}{\partial t_0^P}\Bigl(G_{n,\alpha}(L)\Bigr)_0,\\
\frac{\partial v}{\partial t_n^{\alpha}}&=&
\frac{\partial}{\partial t_0^P}\Bigl(G_{n,\alpha}(L)\Bigr)_{-1},
\eeqa
where $\Bigl(G_{n,\alpha}\Bigr)_0$ and $\Bigl(G_{n,\alpha}\Bigr)_{-1}$
denote the constant (or $p$-independent) term
and the coefficient of $p^{-1}$ of $G_{n,\alpha}$ respectively.
This implies
\beqa
\bra \sigma_n(O_{\alpha})P\ket&=&\Bigl(G_{n,\alpha}(L)\Bigr)_0,\\
\bra \sigma_n(O_{\alpha})Q\ket&=&\Bigl(G_{n,\alpha}(L)\Bigr)_{-1},
\eeqa
or equivalently,
\beq
\bra \sigma_n(O_{\alpha})O_{\beta}\ket=
\oint G_{n,\alpha}(L)\widehat{O}_{\beta}\frac{dp}{p},
\label{sourse}
\eeq
where $\widehat{P}=1$ and $\widehat{Q}=p$.
One can check that (\ref{sourse}) satisfy
the fundamental recursion relation (\ref{recrel}) by using 
(\ref{MCP1}) for the matrix $(M_{\alpha}^{\,\,\beta})$
and hence provide a correct representation of two point functions.

\bigskip
\noindent
{\sc 4.2 Generalization}

\medskip
In the case of a general target space, we search for the Lax operator 
so that the two point functions can be expressed
in a way similar to (\ref{sourse}): Let $L$ be a polynomial
of several variables $X_1,X_2,\ldots,X_{N}$ and their inverse powers,
whose coefficients are given in terms of the order parameters
$u_{\alpha}$. We require that it satisfies
\beq
\bra\sigma_n(O_{\alpha})O_{\beta}\ket
=\oint G_{n,\alpha}(L)\widehat{O}_{\beta}\,\Omega,
\label{nalO}
\eeq
\beq
\Omega=\frac{dX_1}{X_1}\wedge \cdots\wedge\frac{dX_{N}}{X_{N}},
\eeq
for a suitable choice of functions
$G_{n,\alpha}(L)$ and $\widehat{O}_{\beta}(X_i)$.
We assume $\widehat{P}=1$.

\medskip
We first see that we can almost determine the function
$G_{n,\alpha}(L)$ under a few assumptions on $L$.
Note that the operator
\beq
{\cal D}=M_{0\beta}\frac{\partial}{\partial u_{\beta}}
\eeq
counts the dimension assigned to the parameters $u_{\alpha}$
in such a way that
\beqa
&&[u_{\alpha}]=q_{\alpha}+1-\dim M,\\
&&{[}\exp(u^jd_j){]}=c_1(M)\cdot d.
\eeqa
We assume

\vspace{2mm}

\noindent
\underline{Affine Dependence on $u^0=\eta^{0\beta}u_{\beta}$}:
\beq
L=u^0+(\mbox{$u^0$-independent terms}).
\eeq

\noindent
\underline{Homogeneity}:
\beq
{\cal D}L+\sum_i q_i X_i\frac{\partial}{\partial X_i}L=L,
\eeq
where $q_i$ is a dimension assigned to $X_i$.

Recall that the two point functions satisfy for $n\geq 1$
(see lemma 1 and (\ref{recrel}))
\beqa
&&\frac{\partial}{\partial u^0}\bra\sigma_n(O_{\alpha})P\ket
=n\bra\sigma_{n-1}(O_{\alpha})P\ket,
\label{eqtp}\\
&&{\cal D}\bra\sigma_n(O_{\alpha})P\ket
=(n+1+q_{\alpha}-\dim M)\bra\sigma_n(O_{\alpha})P\ket
+n\bra\sigma_{n-1}(c_1(M)\wedge O_{\alpha})P\ket.
\eeqa
Under the above assumptions, these imply
\beqa
\frac{d}{dL}G_{n,\alpha}&=&nG_{n-1,\alpha},
\label{eq1Gna}\\
L\frac{d}{dL}G_{n,\alpha}&=&(n+1+q_{\alpha}-\dim M)G_{n,\alpha}
+nc_1(M)_{\alpha}^{\beta}G_{n-1,\beta}.
\label{eq2Gna}
\eeqa
These equations have enough power to determine
the functions $G_{n,\alpha}$ up to some finite number of
arbitrary constants.
For instance, in the case of the volume class $\Top$ 
the second term of (\ref{eq2Gna})
is absent and we have
(up to an overall constant)
\beq
G_{n,\Top}=\frac{1}{n+1}L^{n+1}.
\eeq
In particular, we can use the equation
\beq
\bra \sigma_n(\Top)P\ket
=\frac{1}{n+1}\oint L^{n+1}\Omega,
\label{defLax}
\eeq
to determine $L$.

\medskip
In the rest of the paper, we focus our attention on the
subspace of the phase space where only the relevant ($t_{0,P}$)
and the marginal (K\"ahler) perturbations are turned on.
We often turn off also the relevant perturbation,
since its dependence is controlled by the puncture equation
(\ref{punc eq}) and is easy to recover.
On this subspace,
the fundamental matrix is greatly simplified
and this enables us to find a simple expression for the Lax operator
for a wide class of target spaces.

The subspace of moduli space where only the marginal (K\"ahler) perturbation 
is turned on has been the area
where the quantum cohomology and mirror symmetry played an important role.
The objective of this paper 
is to extend the theory, on the same subspace,
to the case of {\it gravitational} quantum cohomology where 
the two-dimensional gravity is coupled to the $\sigma$ model and the
Mumford-Morita
classes interact with the cohomology classes of $M$.
As we are going to discuss in the following, we will observe a mirror symmetry
in a generalized sense when the system is coupled to gravity.

\bigskip
\noindent{\sc 4.3 $\CP^2$ Model In Detail}

\medskip
For the case of $\CP^2$ model, the equations
(\ref{eq1Gna}) and (\ref{eq2Gna})
for $G_{n,\alpha}$
are
\beqa
\frac{d}{dL}G_{n,\alpha}&=&nG_{n-1,\alpha},\qquad\alpha=P,Q,R,\\
L\frac{d}{dL}G_{n,R}&=&(n+1)G_{n,R},\\
L\frac{d}{dL}G_{n,Q}&=&n G_{n,Q}+3n G_{n-1,R},\\
L\frac{d}{dL}G_{n,P}&=&(n-1)G_{n,P}+3n G_{n-1,Q}.
\eeqa
The general solution is
\beqa
G_{n,R}&=&\frac{1}{n+1}L^{n+1},
\label{GnR}\\[0.2cm]
G_{n,Q}&=&3 L^n(\log L-c_n+c),
\label{GnQ}\\[0.2cm]
G_{n,P}&=&9nL^{n-1}\left(\,\frac{1}{2}(\log L)^2-c_{n-1}\log L+d_{n-1}
+c(\log L-c_{n-1})+d\,\right),
\label{GnP}
\eeqa
where $c$ and $d$ are arbitrary constants and
\beq
d_n=1+\frac{c_2}{2}+\cdots+\frac{c_n}{n}.
\label{dn}
\eeq
We will find the Lax operator $L$
so that (\ref{nalO}) holds with
these expressions for $G_{n,\alpha}$
(for a suitable choice of $c$ and $d$).

\bigskip
\noindent
{\it The Two Point Functions}

\medskip
When we turn of all the couplings except $t_0^Q=t$,
the fundamental matrix (\ref{MCP2}) is simplified as
\beq
\left(
\begin{array}{ccc}
0&3&0\\
0&0&3\\
\!3\e^t\!\!&0&0
\end{array}
\right).
\eeq
Then, the recursion relation (\ref{recrel}) reads as
\beq
\bra \sigma_n(O_{\alpha})O_{\beta}\ket
=\frac{3n}{(n+\alpha+\beta-1)}\left\{
\begin{array}{cl}
\bra\sigma_{n-1}(O_{\alpha})O_{\beta+1}\ket-
\bra\sigma_{n-1}(O_{\alpha+1})O_{\beta}\ket &O_{\beta}\ne R\\[0.25cm]
\e^t\bra\sigma_{n-1}(O_{\alpha})P\ket-
\bra\sigma_{n-1}(O_{\alpha+1})O_{\beta}\ket &O_{\beta}= R
\end{array}\right.
\label{recrelCP2}
\eeq

\medskip
\noindent
where $O_0=P,O_1=Q,O_2=R$ and $O_3=0$.
Together with the boundary conditions $\bra RR\ket=\e^t$,
$\bra P Q\ket=t$ and $\bra \sigma_1(P)P\ket=t^2/2$, (\ref{recrelCP2})
yields
\beqa
\bra\sigma_{3m-1}(R)P\ket&=&\frac{(3m-1)!}{(m!)^3}\e^{mt},
\label{RP}\\[0.2cm]
\bra\sigma_{3m}(Q)P\ket&=&(t-3c_m)\frac{(3m)!}{(m!)^3}\e^{mt},
\label{QP}\\[0.2cm]
\bra\sigma_{3m+1}(P)P\ket&=&
\left(\frac{t^2}{2}-3c_m t+9d_m-3\widetilde{c_m}\right)
\frac{(3m+1)!}{(m!)^3}\e^{mt},
\label{PP}
\eeqa
where $\widetilde{c_m}=\sum_{j=1}^m1/j^2$.
On dimensional ground,
$\bra \sigma_n(O_{\alpha})P\ket=0$ if
$n+\alpha-1\not\equiv 0$ mod 3.

\bigskip
\noindent
{\it The Lax Operator} 

\medskip
Let us introduce two variables $X$ and $Y$.
The value (\ref{RP}) is nothing but the constant term of
\beq
\frac{1}{3m}(X+X^{-1}Y+\e^tY^{-1})^{3m},
\eeq
There are no constant terms in $(X+X^{-1}Y+\e^tY^{-1})^{3m\pm 1}$.
Thus, we define the Lax operator by

\beq
L=X+X^{-1}Y+\e^tY^{-1}.
\eeq

\medskip
\noindent
It is easy to see that the equation
(\ref{nalO}) for $O_{\alpha}=R$ holds if we take
\beq
\Omega=\frac{dX}{X}\wedge\frac{dY}{Y},
\eeq
and
\beq
\widehat{P}=1,\qquad\widehat{Q}=X,\qquad\widehat{R}=Y.
\eeq
The logarithm of $L$ is defined by
taking the ``average'' of three kinds of expansions:
\beqa
\log L&=&
\frac{1}{3}\left(\,\,\log (1+X^{-2}Y+\e^t X^{-1}Y^{-1})
+\log (X^2 Y^{-1}+1+\e^t X Y^{-2})\right.\nonumber\\
&&\left.
\qquad+t+\log (\e^{-t}XY+\e^{-t}X^{-1}Y^2+1)\,\,\right)\\
=\frac{t}{3}+\frac{1}{3}\sum_{n=1}^{\infty}\!\!\!\!&&\!\!\!\!
\!\!\!\!\frac{(-1)^{n+1}}{n}
\left\{(X^{-2}Y+\e^t X^{-1}Y^{-1})^n
+(X^2 Y^{-1}+\e^t X Y^{-2})^n+(\e^{-t}XY+\e^{-t}X^{-1}Y^2)^n\right\}.
\nonumber
\eeqa
The square of $\log L$ can also be defined, though we need a care for 
cross terms between different types of expansions.
Then, we can check that (\ref{GnQ}),(\ref{GnP})
satisfy the equation (\ref{nalO})
if we take $c=0$ and $d=-\zeta(2)/3=-\sum_{n=1}^{\infty}1/(3n^2)$.
(checked for $G_{n,Q}$, $n=1,\ldots,18$ and
$G_{n,P}$, $n=2,\ldots,7$.) 
To summarize, the expression for the descendants reads as
\beqa
G_{n,R}&=&\frac{1}{n+1}L^{n+1},
\label{nR}\\[0.2cm]
G_{n,Q}&=&3 L^n(\log L-c_n),
\label{nQ}\\[0.2cm]
G_{n,P}&=&9nL^{n-1}\left(\,\frac{1}{2}(\log L)^2-c_{n-1}\log L+d_{n-1}
-\frac{1}{3}\zeta(2)\,\right),
\label{nP}
\eeqa
where $c_n$ and $d_n$ are given in (\ref{cn}) and (\ref{dn}) respectively.

\bigskip
\noindent
{\sc 4.4 The $\CP^N$ Model}

\medskip
It is straightforword to generalize the above results
to the case of $\CP^N$ model.
The $\CP^N$ model has $N+1$ primaries $O_0=P,O_1,\ldots ,O_N$
corresponding to $1,\omega,\ldots,\omega^N\in H^*(\CP^N)$ respectively,
where $\omega$ is the K\"ahler class with a unit volume.
Here we consider only the marginal perturbation $t_0^1=t$.
The Lax operator is expressed in terms of $N$ variables
$X_1,\ldots, X_N$ by
\beq
L=X_1+X_1^{-1}X_2+\cdots +X_{N-1}^{-1}X_N+\e^t X_N^{-1}.
\label{LPN}
\eeq
This satisfies
\beq
\bra \sigma_n(O_{\alpha})O_{\beta}\ket
=\oint G_{n,\alpha}(L)\widehat{O}_{\beta}\,\Omega,
\eeq
\beq
\Omega=\frac{dX_1}{X_1}\wedge \cdots\wedge\frac{dX_{N}}{X_{N}}.
\label{holOm}
\eeq
The primary fields $O_i$ ($i=0,1,\ldots,N$) are represented by
\beq
\widehat{O}_i=X_i.
\eeq
The representation $G_{n,\alpha}$ for descendants are determined
(up to some constants of integration) by the equations (\ref{eq1Gna}) and
(\ref{eq2Gna}). Some of them are given by
\beqa
&&G_{n,N}=\frac{1}{n+1}L^{n+1}
\label{nN}\\
&&G_{n,N-1} = (N+1) L^n(\log L-c_n^{(1)}),
\label{nN-1}\\
&&G_{n,N-2}=(N+1)^2n L^{n-1}
\left(\frac{(\log L)^2}{2}-c_{n-1}^{(1)}\log L+c_{n-1}^{(2)}\right),
\label{nN-2}\\
&&G_{n,N-3}=(N+1)^3n(n-1) L^{n-2}
\left(\frac{(\log L)^3}{3!}-c_{n-2}^{(1)}\frac{(\log L)^2}{2}
+c_{n-2}^{(2)}\log L-c_{n-2}^{(3)}\right),
\label{nN-3}\nonumber\\[-0.1cm]
&& \null \\[-0.4cm]
&&\hspace{1cm}\cdots\cdots
\nonumber
\eeqa
where $c^{(i)}_n$ are constants satisfying
$c_n^{(i)}=c_{n-1}^{(i)}+c_{n}^{(i-1)}/n$
and $c^{(1)}_n=c_n$.

\bigskip
\noindent
{\sc 4.5 Landau-Ginzburg Formulation}

\medskip
In the tree-level topological string theory
of minimal \cite{DVV,EKYY} or $\CP^1$ model \cite{EHY},
a Landau-Ginzburg (LG) description has been developped
where the superpotential is given by the Lax operator of the
corresponding integrable system.
Here, we develop a LG formulation for the $\CP^N$ model.

\medskip
The superpotential we consider here is the Lax operator (\ref{LPN}).
When the relevant perturbation ($t_0^P$)
is added, it is given by
\beq
L=t_0^P+X_1+X_1^{-1}X_2+\cdots+X_{N-1}^{-1}X_N+\e^t X_N^{-1}.
\label{LPN2}
\eeq
Topological LG model \cite{Vafa}
is a B-twisted $N=2$ sigma model (equipped with
the F-term potential $\int d^2\theta L$) and
upon quantization a nowhere vanishing
holomorphic form of middle dimension must be specified \cite{W3}.
In particular, the target space must be a Calabi-Yau manifold.
In our case, we take as the target space the algebraic torus
$(\C^*)^N$ with the holomorphic $N$-form (\ref{holOm}):
\beq
\Omega=\frac{dX_1}{X_1}\wedge \cdots\wedge\frac{dX_{N}}{X_{N}}.
\eeq
The vacua are identified with the critical points of $L$,
$X_i\partial_{X_i}L=0$:
\beq
X_1=X_1^{-1}X_2=X_2^{-1}X_3=\cdots =X_{N-1}^{-1}X_N=\e^t X_N^{-1},
\eeq
which consist of the $N+1$ points
\beqa
X_{*1}&=&
\e^{\frac{t}{N+1}}\zeta^k, \hspace{3mm} k=0,1,2\cdots N, \\
X_{*j}&=&(X_{*1})^j, \hskip2mm j=2,\cdots,N
\eeqa
where $\zeta=\exp(\frac{2\pi i}{N+1})$.
These are all non-degenerate (Hessians are non-vanishing)
and hence the number of vacua
is $N+1$ which is consistent with ${\rm Tr}(-1)^F=\chi(\CP^N)=N+1$
of the $\CP^N$ sigma model.

The three point function is given by
\beqa
\bra A B C\ket&=&
\sum_{X_*:\,{\rm critical}}
\frac{A(X_*)B(X_*)C(X_*)}{{\rm Hess}_{X_*}(L)}
\label{Res1}\\[0.2cm]
&=&\oint\frac{A(X)B(X)C(X)}{\prod_{j=1}^NX_j\partial_{X_j}L}\,\Omega,
\label{Res2}
\eeqa
where the last integration is over the small $N$-dimensional
tori encircling the
$N+1$ vacua \cite{GH}.
The Hessian ${\rm Hess}_{X_*}(L)$ at the critical point $X_*$
is given by
\newfont{\bg}{cmr10 scaled\magstep4}
\newcommand{\bigzerol}{\smash{\hbox{\bg 0}}}
\newcommand{\bigzerou}{\smash{\lower1.7ex\hbox{\bg 0}}}
\beqa
{\rm Hess}_{X_*}(L)&:=&\det(X_i\partial_{X_i}X_j\partial_{X_j}L)|_{X=X_*}\\
&=&(X_{*1})^{\,\,N}\left|\begin{array}{cccccc}
\,2&\!\!\!-1&&&&\!\!\!\bigzerou\,\,\,\,\\[0.2cm]
\,\!\!\!-1&2&\!\!\!-1&&&\\
&\!\!\!-1&2&\ddots&&\\
&&\ddots&\ddots&\ddots&\\
&&&\ddots&2&\!\!\!-1\\[0.2cm]
\,\,\,\,\,\bigzerol\!\!&&&&\!\!\!-1&2
\end{array}
\right|\,\,=\,\,(N+1)(X_{*1})^{\,\,N}.
\eeqa
Thus, for example,
\beqa
\bra P X_i X_j\ket&=&\sum_{\rm vacua}
\frac{(X_{*1})^{\,\,i+j}}{(N+1)(X_{*1})^{\,\,N}}\,=\,\delta_{i+j,N}\\[0.1cm]
\bra X_1 X_i X_j\ket&=&\sum_{\rm vacua}
\frac{(X_{*1})^{\,\,i+j+1}}{(N+1)(X_{*1})^{\,\,N}}\,=\,
\delta_{i+j+1,N}+\e^t\delta_{i,N}\delta_{i,N}.
\eeqa
These coincide with the 3-point functions $\bra P O_iO_j\ket$
and $\bra O_1O_iO_j\ket$, respectively, of the $\CP^N$ model
and reproduce the quantum cohomology relation
\beq
X_1^{N+1}=\e^t.
\eeq
Therefore, we can identify $\widehat{O}_i=X_i$
as the LG representative for the
primary field $O_i$ of the $\CP^N$ model.

\medskip
Next, we provide a LG description for gravitational descendants
$\sigma_n(O_i)$. Here, it is convenient to introduce new LG variables
$Z_1,\ldots.Z_N$ defined by
\beq
Z_1=X_1,\quad Z_2=X_1^{-1}X_2,\quad Z_3=X_2^{-1}X_3,\quad\ldots\quad,
Z_N=X_{N-1}^{-1}X_N.
\eeq
The holomorphic form $\Omega$ is still represented as
$
\Omega=\frac{dZ_1}{Z_1}\wedge\cdots\wedge\frac{dZ_N}{Z_N}
$
and therefore, the residue formulae (\ref{Res1}),(\ref{Res2})
also holds with this choice of variable.
The Lax operator is expressed as
$L=t_0^P+Z_1+Z_2+\cdots+Z_N+\e^tZ_1^{-1}\cdots Z_N^{-1}$.
The three point function
$\bra \sigma_n(O_i)PP\ket$ is given by
\beq
\bra\sigma_n(O_i)PP\ket
=\frac{\partial}{\partial t_{0,P}}\bra\sigma_n(O_i)P\ket
=\oint G_{n,i}^{\prime}(L)\,\Omega,
\eeq
where $G_{n,i}^{\prime}(L)=dG_{n,i}/dL$ and
the integration is performed along
$|Z_i|={\rm const}$.
The contours can be deformed to large ones
$|Z_i|={\rm const}\gg 1$
and then we find
\beq
\bra\sigma_n(O_i)PP\ket
=\oint \,
\frac{\widehat{\sigma_n}(O_i)}{\prod_{j=1}^NZ_j\partial_{Z_j}L}\,\Omega,
\label{Res3}
\eeq
where
\beq
\widehat{\sigma_n}(O_i)
=\biggl(G_{n,i}^{\prime}(L)\prod_{j=1}^N
Z_j\partial_{Z_j}L\biggr)_{\! +}.
\label{LGdes}
\eeq
Here $(\cdots)_+$ is the projection to non-negative powers of $Z_i$:
\beq
\Bigl(Z_1^{n_1}\cdots Z_N^{n_N}\Bigr)_+\,\,:=\,\left\{
\begin{array}{cl}
Z_1^{n_1}\cdots Z_N^{n_N}&\mbox{if $n_1,\ldots,n_N\geq 0$}\\
0&\mbox{otherwise.}
\end{array}
\right.
\label{def+}
\eeq
We propose (\ref{LGdes})
as the LG representative for the descendants.
This identification may be justified by the fact that (\ref{LGdes})
satisfies the ``topological recursion relation'':
\beq
\widehat{\sigma_n}(O_i)\,\equiv\,
n\,\bra\sigma_{n-1}(O_i)O^j\ket\widehat{O}_j,
\eeq
modulo terms that vanish at the critical points
(BRST-exact terms divisible by
$Z_j\partial_{Z_j}L=Z_j-\e^tZ_1^{-1}\cdots Z_N^{-1}$ for some $j$).
Note that the fields $\widehat{O}_j$ are given by
\beq
\widehat{O}_j=X_j=Z_1\cdots Z_j\equiv Z_{i_1}\cdots Z_{i_j}
\eeq
for arbitrary $i_1,\ldots,i_j$, and also
\beq
n\,\bra\sigma_{n-1}(O_i)O_j\ket
=\oint G_{n,i}^{\prime}(L)X_j\Omega
=\left(\,G_{n,i}^{\prime}(L)Z_{i_1}\cdots Z_{i_j}\,\right)_0
\eeq
for {\it distinct} $i_1,\ldots,i_j$,
where $(\cdots)_0$ denotes the term independent of $Z_1,\ldots,Z_N$.
Therefore, what we need to establish is
\beqa
\biggl(f(L)\prod_{j=1}^N
Z_j\partial_{Z_j}L\biggr)_{\! +}&\equiv&
\Bigl(f(L)\Bigr)_0Z_1\cdots Z_N+\Bigl(f(L)Z_1\Bigr)_0Z_2\cdots Z_N+\cdots
\label{LGtrr}\\
&&
\hspace{0.5cm}
\cdots +\Bigl(f(L)Z_1\cdots Z_{N-1}\Bigr)_0Z_N
+\Bigl(f(L)Z_1\cdots Z_N\Bigr)_0,
\nonumber
\eeqa
for an arbitrary function $f(L)$ of $L$.
This is the multi-variable version of a similar formula
for the one-variable case of minimal models \cite{EKYY}
and the $\CP^1$ model \cite{EHY}.
In Appendix A, we present a proof of (\ref{LGtrr}) in the case of 
$\CP^2$ model.

The LG representative (\ref{LGdes})
also satisfies an equation
\beq
\frac{\partial}{\partial t_{0,P}}\widehat{\sigma_n}(O_i)
=\biggl(G_{n,i}^{\prime\prime}(L)\prod_{j=1}^N
Z_j\partial_{Z_j}L\biggr)_{\! +}
=n\widehat{\sigma_{n-1}}(O_i)
\label{LGpunc}
\eeq
as a consequence of the $t_{0,P}$-dependence of
$L=t_{0,P}+\cdots$ and the identity
$G_{n,i}^{\prime}(L)=nG_{n-1,i}(L)$.
(\ref{LGpunc}) leads to the puncture equation (\ref{punc eq}).
Note that the shift of the potential
by the ``cosmological constant'' $t_{0,P}$
is irrelevant as far as we only consider 
primary three-point functions.
This is the case for a theory without gravity.
However, 
the shift has a non-trivial effect in the three point functions
including the descendants.
This is a manifestation of the
effect of two-dimensional gravity.

The equation (\ref{kahler eq}) for the K\"ahler class can also
be derived in the LG description
in the same way as in the $\CP^1$ model \cite{EHY}.
We turn off $t_{0,P}$.
Let us introduce new LG variables $\overline{Z}_i=\e^{-\frac{t}{N+1}}Z_i$.
Then, $\overline{L}=\e^{-\frac{t}{N+1}}L$ has an expression
independent of $t$,
and the $t$ dependence in
\beq
\overline{\sigma_n}(O_i):=\e^{-\frac{n+i}{N+1}t}\widehat{\sigma_n}(O_i)
\eeq
appears only through logarithms.
Then, it is easy to see
\beq
\frac{\partial}{\partial t}\overline{\sigma_n}(O_i)
=n\overline{\sigma_{n-1}}(O_{i+1})
\eeq
($O_{N+1}:=0$), which leads to (\ref{kahler eq}).

\bigskip
The superpotential (\ref{LPN2}) is nothing but the potential of
the affine Toda field theory of ${\rm A}_N$ type.
Thus, this generalizes the correspondence of
$\CP^1$ model and sine-Gordon theory found in the
previous work \cite{EHY}. The relation of the $\CP^N$ sigma model
and the ${\rm A}_N$ affine Toda field theory has been observed
at various stages. (For $N=2$ supersymmetric models, see
\cite{FI} for the comparison of the S-matrices and
\cite{CV1,CV2} for the $t$-$t^*$ equation
obeyed by the ground state metric.
There exist also $N=0$ literature \cite{Fateev,Zitni}.)
More close to our situation is the work of Batyrev \cite{Baty}.
He observed that the quantum cohomology ring of toric manifolds
(including $\CP^N$) is the Jacobian ring of the superpotentials
of the affine Toda type.
In the case of $\CP^N$, our result shows that
the description in terms of the affine Toda potential extends
to the gravitational quantum cohomology.

\medskip
In the next section, we treat several examples of Fano manifolds $M$.
In each of them,
we find a Lax operator whose number of variables is equal to
the dimension of the original manifold.
It turns out that when $M$ is a toric manifold,
the LG description is given based on
the algebraic torus $(\C^{*})^{\dim M}$.
When $M$ is not toric (e.g. Grassmannians),
the LG desciption on $(\C^{*})^{\dim M}$
has a possible trouble due to a run-away behaviour of the potential.
In such a case, however, the disease may be cured 
if we partially compactify $(\C^{*})^{\dim M}$.
We may refer to the correspondence of the
topological string model based on a Fano manifold
and the LG model based on a non-compact CY manifold
with superpotential of the affine Toda type
as a mirror symmetry in a generalized sense.
Here deformation of the K\"ahler class in the sigma model
side corresponds to deformation of superpotential in the LG.
In the case of ordinary quantum cohomology 
the LG variables are identified as the generators of the
classical cohomology ring $H^*(M)$ and do not possess direct
geometrical significance.
In the gravitational cohomology, however, we have 
LG variables as many as $\dim M$ and they are
identified as the coordinates of manifolds
of the type of algebraic torus $(\C^{*})^{\dim M}$.

\vspace{0.7cm}
\renewcommand{\theequation}{5.\arabic{equation}}\setcounter{equation}{0}

\noindent{\large \bf 5. Examples}

\bigskip
In this section, we construct Lax operators
for more general target spaces. We turn off all the couplings except
for the marginal (K\"ahler) perturbation.
We recall from \S 4.2 (see (\ref{defLax}))
that the Lax operator $L$
for a theory with target space $M$ is defined as a polynomial
in several variables $X_1^{\pm1},\ldots ,X_m^{\pm 1}$
such that
\beq
\bra \sigma_n(\Top)O_{\alpha}\ket
=\frac{1}{n+1}\oint L^{n+1}\,\widehat{O}_{\alpha}\,
\frac{dX_1}{X_1}\wedge \cdots\wedge\frac{dX_{m}}{X_{m}},
\label{defLax5}
\eeq
hold for a suitable expression $\widehat{O}_{\alpha}$,
where $\Top$ is the volume class of $M$.
We will consider (i) product spaces,
(ii) rational surfaces, and we also treat 
(iii) Grassmannians in Appendix B.

\bigskip
\noindent
{\sc 5.1 Product Spaces}

\medskip
Let $M$ and $N$ be two K\"ahler manifolds with positive
first Chern classes.
Suppose that we already have the Lax operators $L_M$ and
$L_N$ of the topological string theories on $M$ and $N$.
Then, we will see under a plausible assumption that
{\it the Lax operator of
the theory for $M\times N$ is given by}
\medskip
\beq
L_{M\times N}=L_M+L_N.
\eeq

\medskip
Let $\{\Omega_{\alpha}\}$ and $\{\omega_j\}$ be basis of the cohomology
groups $H^*(M)$ and $H^*(N)$ respectively,
where $\Omega_{\alpha}$ and $\omega_{j}$ have dimensions $q_{\alpha}$
and $q_j$. We denote by $O_{\alpha},O_j$ and $O_{\alpha,j}$
the primary fields for the classes $\Omega_{\alpha},\omega_{j}$
and $\Omega_{\alpha}\wedge \omega_j$.
One of the most important fact in quantum cohomology of a product is that
the 3-point functions factorize:
\beq
\bra O_{\alpha,i}O_{\beta,j}O_{\gamma,k}\ket
=\bra O_{\alpha}O_{\beta}O_{\gamma}\ket
\bra O_i O_j O_k\ket.
\label{factorize}
\eeq
This can be seen by noting that the evaluation maps
commute with the map
\beq
\overline{\cal M}_{0,3}(M\!\times\! N,d_M\oplus d_N)
\longrightarrow
\overline{\cal M}_{0,3}(M,d_M)\times 
\overline{\cal M}_{0,3}(N,d_N),
\eeq
between moduli spaces of stable maps, which is isomorphic among
open dense subsets corresponding to smooth curves.
(See also \S 2.5 of \cite{KM}, and \cite{KM2}.)
Since the first Chern class of $M\times N$ is just a sum of those for
$M$ and $N$, $c_1(M\times N)=c_1(M)+c_1(N)$,
we see from (\ref{factorize}) that 
the fundamental matrix for $M\times N$ is given by
\beq
M_{\alpha,i}^{\,\,\,\beta,j}\,=\,
M_{\alpha}^{\,\,\beta}\delta_i^{\,j}
+\delta_{\alpha}^{\,\beta}N_i^{\,\,j},
\label{matrixMN}
\eeq
where $M_{\alpha}^{\,\,\beta}$ and $N_i^{\,\,j}$
are the fundamental matrices for $M$ and $N$ respectively.
As a consequence, the two point functions of primaries decompose as
\beq
\bra O_{\alpha,i}O_{\beta,j}\ket
=\bra O_{\alpha}O_{\beta}\ket\,\eta_{ij}
+\eta_{\alpha\beta}\,\bra O_iO_j\ket.
\label{decom}
\eeq
By our assumption,
the Lax operators $L_M$ and $L_N$ are defined as
polynomials in several variables, say $Z_1,\ldots,Z_{\mu}$ for $L_M$
and $W_1,\ldots,W_{\nu}$ for $L_N$ so that the following formulae hold
\beqa
\bra \sigma_n(\Top)O_{\alpha}\ket &=&
\frac{1}{n+1}\Bigl(L_M^{n+1}\widehat{O}_{\alpha}\Bigr)_{\!0},
\label{M2}\\
\bra \sigma_n(\TopN)O_i\ket &=&
\frac{1}{n+1}\Bigl(L_N^{n+1}\widehat{O}_i\Bigr)_{\!0}.
\label{N2}
\eeqa
Here $\widehat{O}_{\alpha}$ and $\widehat{O}_i$ depend on
$Z_a$ and $W_b$, respectively, and $(\cdots)_0$
stands for constant terms, i.e. terms independent of
$Z_a$ or $W_b$.
Now we claim that
\beqa
\bra\sigma_n(\TopMN)O_{\alpha,i}\ket
&=&\frac{1}{n+1}\Bigl((L_M+L_N)^{n+1}\widehat{O}_{\alpha,i}\Bigr)_{\!0,0},
\label{Claim}\\[0.2cm]
&&\hspace{-1cm}\widehat{O}_{\alpha,i}:=\widehat{O}_{\alpha}\widehat{O}_i.
\eeqa
For $n=0$, since $\TopMN=\Top\wedge\TopN$, this follows
from (\ref{decom}) if
\beq
\Bigl(\widehat{O}_{\alpha}\Bigr)_{\!0}=\delta_{\alpha}^{\,0},\qquad
\Bigl(\widehat{O}_i\Bigr)_{\!0}=\delta_i^{\,0},
\eeq
where $\alpha=0$ (or $i=0$) stands for the identity class.
We also assume that these hold.
It is then a straightforward calculation to see that the right hand side of
(\ref{Claim}) satisfies the fundamental recursion relation (\ref{recrel1})
for the theory of $M\times N$ with the matrix being given by
(\ref{matrixMN}):
\beqa
\lefteqn{
(n+1+q_{\alpha}+q_i)
\frac{1}{n+1}
\Bigl((L_M+L_N)^{n+1}\widehat{O}_{\alpha}\widehat{O}_i\Bigr)_{\!0,0}
}\nonumber\\
&&
=(n+1+q_{\alpha}+q_i)\frac{1}{n+1}
\sum_{l=0}^{n+1}{n+1\choose{l}}
\Bigl(L_M^l\widehat{O}_{\alpha}\Bigr)_{\!0}
\Bigl(L_N^{n+1-l}\widehat{O}_i\Bigr)_{\!0}
\nonumber\\
&&=\frac{1}{n+1}\sum_{l=0}^{n+1}{n+1\choose{l}}
\left\{\,l\,\frac{l+q_{\alpha}}{l}+(n+1-l)\frac{n+1-l+q_i}{n+1-l}\,\right\}
\Bigl(L_M^l\widehat{O}_{\alpha}\Bigr)_{\!0}
\Bigl(L_N^{n+1-l}\widehat{O}_i\Bigr)_{\!0}
\nonumber\\
&&
=\frac{1}{n+1}\sum_{l=0}^{n+1}{n+1\choose{l}}
\left\{\,lM_{\alpha}^{\,\,\beta}
\Bigl(L_M^{l-1}\widehat{O}_{\beta}\Bigr)_{\!0}
\Bigl(L_N^{n+1-l}\widehat{O}_i\Bigr)_{\!0}
+(n+1-l)N_i^{\,\,j}
\Bigl(L_M^l\widehat{O}_{\alpha}\Bigr)_{\!0}
\Bigl(L_N^{n-l}\widehat{O}_j\Bigr)_{\!0}
\right\}
\nonumber\\
&&
=\sum_{l=1}^{n+1}{n\choose{l-1}}M_{\alpha}^{\,\,\beta}
\Bigl(L_M^{l-1}\widehat{O}_{\beta}\Bigr)_{\!0}
\Bigl(L_N^{n+1-l}\widehat{O}_i\Bigr)_{\!0}
+\sum_{l=0}^n{n\choose{l}}
N_i^{\,\,j}
\Bigl(L_M^l\widehat{O}_{\alpha}\Bigr)_{\!0}
\Bigl(L_N^{n-l}\widehat{O}_j\Bigr)_{\!0}
\nonumber\\[0.2cm]
&&
=
M_{\alpha}^{\,\,\beta}
\Bigl((L_M+L_N)^n\widehat{O}_{\beta}\widehat{O}_i\Bigr)_{\!0,0}
+N_i^{\,\,j}
\Bigl((L_M+L_N)^n\widehat{O}_{\alpha}\widehat{O}_j\Bigr)_{\!0,0}.
\eeqa
In the third step, we have used the fundamental recursion relation
for each theory.
Since the initial value at $n=0$ and the recursion relation
completely determine the two point functions,
the claim (\ref{Claim}) holds for all $n$.
Thus, we can identify $L_M+L_N$ as the Lax operator for the product 
space $M\times N$.
Note that the number of variables of $L_{M+N}$ equals
dim$M$+dim$N$ which
coincides again with the dimension of $M\times N$.
We also note that we recover the relation
$\chi(M\times N)=\chi(M)\chi(N)$
for the supersymmetry index,
since the number of vacua equals that of the critical points
of the superpotential in the LG description: the
number of critical points of $L_{M+N}$ is given by the
product of those of $L_{M}$ and $L_{N}$.

Suppose that the gravitational descendants
are representable as LG fields in theories for $M$ and $N$.
Namely, we assume for a suitable choice of LG variables $Z_a$
and $W_b$, we have
\beqa
\Bigl(f(L_M)\prod_aZ_a\partial_{Z_a}L_M\Bigr)_{\!+}
&\equiv&
\Bigl(f(L_M)\widehat{O}^{\alpha}\Bigr)_{\!0}\widehat{O}_{\alpha},\\
\Bigl(f(L_N)\prod_bW_b\partial_{W_b}L_N\Bigr)_{\!+}
&\equiv&
\Bigl(f(L_N)\widehat{O}^i\Bigr)_{\!0}\widehat{O}_i,
\eeqa
modulo terms that vanish at the critical points.
Here $(\cdots)_+$ are defined as in (\ref{def+}).
Then, we find
\beqa
\Bigl(f(L_M+L_N)\prod_aZ_a\partial_{Z_a}L_M
\prod_bW_b\partial_{W_b}L_N\Bigr)_{\!+,+}
&\equiv&
\Bigl(\Bigl(f(L_M+L_N)\widehat{O}^{\alpha}\Bigr)_{\!0}\widehat{O}_{\alpha}
\prod_bW_b\partial_{W_b}L_N\Bigr)_{\!+}\nonumber\\
&\equiv&
\Bigl(f(L_M+L_N)\widehat{O}^{\alpha}\widehat{O}^i\Bigr)_{\!0,0}
\widehat{O}_{\alpha}\widehat{O}_i.
\eeqa
Thus the descendants of the product theory are also 
representable as LG fields.

\medskip
\bigskip
\noindent
{\sc 5.2 Rational Surfaces}

\medskip
It is interesting to see how the Lax operator behaves under
birational transformations of the target space.
The simplest example for such a study is provided by
rational surfaces  -- two dimensional complex manifolds that
can be obtained from $\CP^2$ by a series of blow ups and downs.
There are only ten rational surfaces of positive first Chern classes
(Del Pezzo surfaces, see \cite{Manin}):
$\CP^2$, the quadric $\CP^1\!\times\!\CP^1$, and blow up $M_r$ of
$\CP^2$ at $r$-points in general position ($r=1,\ldots ,8$).
Here we consider $\CP^1\!\times\!\CP^1$, $M_1$ and $M_2$.

In general, the genus $0$ free energy for a complex surface $M$
has the instanton expansion
\beq
F_0=\sum_d \,N_d\,\frac{t_R^{c_1(M)\cdot d-1}}{(c_1(M)\cdot d-1)!}
\,\e^{t\cdot d},
\eeq
where $t_R=t_0^R$ is the coupling constant of the unique irrelevant
operator $R$ (the volume class of $M$).
The sum runs over effective classes, i.e. homology classes
that are positive on the K\"ahler cone.
For Del Pezzo surfaces, the numbers $N_d$ are determined
\cite{KM,I} by the associativity equation
and an initial condition (see also \cite{CM}).
For the space $M_r$, the initial condition
is stated as $N_d=1$ for every exceptional class $d$. (A (co)homology
class $E$ is called exceptional when it
is represented by a $\CP^1$ with a self-intersection number $-1$,
namely, $c_1(M)\cdot E=1$ and $E\cdot E=-1$.)
For our purpose of determining the fundamental matrix,
it is enough to know the free energy $F$ up to order $t_R^2$.

\bigskip
\noindent
{\it The Quadric Surface $\CP^1\times \CP^1$}

\medskip
Since this is a product,
the general argument of the previous subsection should apply.
There are four primaries; $P,Q_1,Q_2,R$ where
$Q_1$ and $Q_2$ correspond to the K\"ahler classes of one $\CP^1$
and the other.
The first Chern class is $2 Q_1+2 Q_2$ and the free energy is given by
\beq
F_0\,\,=
\sum_{d_1,d_2\geq 0}N_{d_1,d_2}\frac{t_R^{2d_1+2d_2-1}}{(2d_1+2d_2-1)!}
\,{\rm e}^{d_1t^1+d_2t^2}
=\,\,
t_R\e^{t^1}+t_R\e^{t^2}+\cdots.
\eeq
From this expansion, we can read off the fundamental matrix
and find that it is indeed a tensor product of
two matrices for $\CP^1$'s:
\beq
\left(
\begin{array}{cccc}
0&2&2&0\\
2\e^{t^1}&0&0&2\\
2\e^{t^2}&0&0&2\\
0&2\e^{t^2}&2\e^{t^1}&0
\end{array}
\right)
=
\left(
\begin{array}{cc}
0&2\\
2\e^{t^1}&0
\end{array}
\right)
\otimes
\left(
\begin{array}{cc}
1&0\\
0&1
\end{array}
\right)
+
\left(
\begin{array}{cc}
1&0\\
0&1
\end{array}
\right)
\otimes
\left(
\begin{array}{cc}
0&2\\
2\e^{t^2}&0
\end{array}
\right).
\eeq
As we have seen in the previous subsection,
the Lax operator is the sum of those 
for the $\CP^1$ model
and the fields are products:
\beq
L=p_1+\e^{t^1}p_1^{-1}+p_2+\e^{t^2}p_2^{-1},
\label{Laxquad}
\eeq
\beq
\widehat{P}=1,\quad
\widehat{Q_1}=p_1,\quad
\widehat{Q_2}=p_2,\quad
\widehat{R}=p_1p_2.
\label{fieldsquad}
\eeq
The theory admits the LG description (based on $\C^*\times\C^*$)
since the $\CP^1$ model does.

\bigskip
\noindent
{\it Blow Up at One Point}

\medskip
The space $M_1$ is obtained by blowing up $\CP^2$ at one point $q\in\CP^2$.
There are four primaries $P,H,E,R$ where
$H$ is the pull back of the K\"ahler class of $\CP^2$ by
the projection $\pi:M_1\to \CP^2$, and $E$ is the unique exceptional class
$[\pi^{-1}(q)]$. The metric is given by
$\eta_{PR}=\eta_{HH}=-\eta_{EE}=1$.
The first Chern class is $3H-E$ and the free energy is given by
($t_0^H=t_H,t_0^E=t_E$)
\beq
F_0\,\,=\sum_{a\geq b,0}N_{a,b}\frac{t_R^{3a-b-1}}{(3a-b-1)!}
\,\e^{at_H+bt_E}
=\,\,\e^{-t_E}+t_R\,\e^{t_H+t_E}+\frac{t_R^2}{2}\,\e^{t_H}+\cdots,
\eeq
which yields the following expression for the fundamental matrix
($\beta_H:=\e^{t_H}$,$\beta_E:=\e^{t_E}$):
\beq
\left(
\begin{array}{cccc}
0&3&-1&0\\
2\beta_H\beta_E&0&0&3\\
2\beta_H\beta_E&0&-\beta_E^{-1}&1\\
3\beta_H&2\beta_H\beta_E&-2\beta_H\beta_E&0
\end{array}
\right).
\eeq
The two point functions are recursively determined
by using this expression.
Then, we can construct the Lax operator and the expression for the primary
fields so that the equation (\ref{defLax5}) holds. The result
is ((\ref{defLax5}) is checked up to $n=15$)
\beqa
L&=&X+X^{-1}Y+\beta_HY^{-1}+\beta_EY,\\[0.2cm]
&&\widehat{P}=1\\
&&\widehat{H}=X+\beta_E Y\\
&&\widehat{E}=\beta_E Y\\
&&\widehat{R}=Y+\beta_E XY.
\eeqa
As $\beta_E\to 0$ (the limit in which the volume of the exceptional curve
$\pi^{-1}(q)$
becomes {\it minus} infinity), $\widehat{E}$ vanishes,
$L$ approaches the Lax operator of $\CP^2$ and $\widehat{P},\widehat{H},
\widehat{R}$ become $\widehat{P},\widehat{Q},\widehat{R}$
of $\CP^2$. We note that $\widehat{E}/\beta_E$ turns into $\widehat{R}$.

\medskip
It is easy to see that the vacuum equation $X\partial_XL=Y\partial_YL=0$
has four non-zero solutions and there is no vacuum at infinity.
This is consistent with the mass gap and the supersymmetry index
$\chi(M_1)=4$ of the sigma model.
Thus, we have a sound LG descrition
with superpotential $L$
based on the algebraic torus
$\C^*\times\C^*=\{(X,Y)\}$.

\newcommand{\beH}{\beta_H}
\newcommand{\beone}{\beta_{E_1}}
\newcommand{\betwo}{\beta_{E_2}}
\newcommand{\bei}{\beta_{E_i}}
\newcommand{\tilY}{\tilde{Y}}
\bigskip
\noindent
{\it Blow Up at Two Points}

\medskip
The space $M_2$ is obtained by blowing up $\CP^2$ at two
points $q_1,q_2\in\CP^2$.
There are five primaries $P,H,E_1,E_2,R$
where $H$ is the pull back of the K\"ahler class of $\CP^2$ by
$\pi_2:M_2\to\CP^2$, and $E_i$ is the exceptional class
$[\pi_2^{-1}(q_i)],\hskip1mm i=1,2$.
There is in addititon another exceptional class $E_{12}=H-E_1-E_2$.
The metric is given by
$\eta_{PR}=\eta_{HH}=-\eta_{E_1E_1}=-\eta_{E_2E_2}=1$.
The first Chern class is $3H-E_1-E_2$ and the free energy is given by
($t_0^H=t_H,t_0^{E_i}=t_{E_i}$)
\beqa
F_0&=&\sum_{a\geq b_1,b_2,0}
N_{a,b_1,b_2}\frac{t_R^{3a-b_1-b_2-1}}{(3a-b_1-b_2-1)!}
\,\e^{at_H+b_1t_{E_1}+b_2t_{E_2}}\\
&=&\e^{t_H+t_{E_1}+t_{E_2}}+\e^{-t_{E_1}}+\e^{-t_{E_2}}
+t_R\,(\e^{t_H+t_{E_1}}+\e^{t_H+t_{E_2}})+\frac{t_R^2}{2}\,\e^{t_H}+\cdots,
\eeqa
which yields the following expression for the fundamental matrix
($\beH:=\e^{t_H}$,
$\bei:=\e^{t_{E_i}}$):
\beq
\left(
\begin{array}{ccccc}
0&3&-1&-1&0\\
2(\beH\beone+\beH\betwo)&\beH\beone\betwo&
-\beH\beone\betwo&-\beH\beone\betwo&3\\
2\beH\beone&\beH\beone\betwo&
-\beone^{-1}-\beH\beone\betwo&-\beH\beone\betwo&1\\
2\beH\betwo&\beH\beone\betwo&
-\beH\beone\betwo&-\betwo^{-1}-\beH\beone\betwo&1\\
3\beH&2(\beH\beone+\beH\betwo)&-2\beH\beone&-2\beH\betwo&0
\end{array}
\right)
\eeq
As in the previous example we can determine the Lax operator
and the representatives for the fields.
The result is ((\ref{defLax5}) is checked up to $n=15$)
\beqa
L&=&X+X^{-1}Y+\beH Y^{-1}+(\beone+\betwo)Y+\beone\betwo XY,
\label{Lax2pt}\\[0.2cm]
&&\widehat{P}=1\\
&&\widehat{H}=X+(\beone+\betwo)Y+2\beone\betwo XY\\
&&\widehat{E}_1=\beone Y+\beone\betwo XY\\
&&\widehat{E}_2=\betwo Y+\beone\betwo XY\\
&&\widehat{R}=Y+(\beone+\betwo)XY+\beone\betwo(2Y^2+(\beone+\betwo)XY^2).
\eeqa
As $\betwo\to 0$,
the Lax operator $L$ and the fields $\widehat{P},\widehat{H},
\widehat{E}_1,\widehat{R}$
become the Lax operator and the fields $\widehat{P},\widehat{H},
\widehat{E},\widehat{R}$ of $M_1$,
while $\widehat{E}_2/\beta_{E_2}\to\widehat{R}$.
Note that the above Lax operator (\ref{Lax2pt}) is manifestly symmetric
under exchange of $\beone$ and $\betwo$.

\medskip
Let us discuss whether a consistent LG description is possible by regarding
(\ref{Lax2pt}) as the superpotential.
Eliminating the variable $X$ from the vacuum equation
$X\partial_XL=Y\partial_YL=0$,
we have
\beqa
&&\beone\betwo(\beone-\betwo)^2Y^5+(\beone-\betwo)^2Y^4
+(2\beH\beone\betwo(\beone+\betwo)-1)Y^3\nonumber\\
&&
+2\beH(\beone+\betwo)Y^2+\beH^2\beone\betwo Y+\beH^2=0
\label{5thorder}
\eeqa
Generically, there are five non-zero solutions to (\ref{5thorder})
and there is no vacuum at infinity,
which is consistent with $\chi(M_2)=5$ and the mass gap.
However, as the K\"ahler structure approaches a $\Z_2$ orbifold point
$\beone=\betwo$ of the moduli space, two of the vacua run away to
infinity and the above LG description breaks down.
Thus, at this point
we must resort to another description that is insensitive
to the singularity of the moduli space.
Let us change the variables as $X\to X$ and $Y\to\tilY=(1+\betwo X)Y$.
Then, we have a new representation
\beqa
L&=&X+X^{-1}\tilY+\beH \tilY^{-1}+\beone\tilY+\beH\betwo X\tilY^{-1},
\label{Lax2pt2}\\[0.2cm]
&&\widehat{H}=X+\beone\tilY+\beH\betwo X\tilY^{-1}\\
&&\widehat{E}_1=\beone \tilY\\
&&\widehat{E}_2=\beH\betwo X\tilY^{-1}\\
&&\widehat{R}=\tilY+\beone X^{-1}\tilY^2.
\eeqa
At the price of losing the manifest $\Z_2$ symmetry,
this description has a good behavior at the orbifold points.
The vacuum equation
$X\partial_XL=\tilY\partial_{\tilY}L=0$ has always
five non-zero solutions and there is no vacuum at infinity
even if $\beone=\betwo$.
Thus, we expect that (\ref{Lax2pt2}) gives a sound LG description
based on $\C^*\times\C^*=\{(X,\tilY)\}$
everywhere on the moduli space of K\"ahler structure.
Away from the orbifold points $\beone=\betwo$,
this is equivalent to the symmetric description (\ref{Lax2pt})
based on $\C^*\times\C^*=\{(X,Y)\}$.

\newcommand{\beQo}{\beta_{Q_1}}
\newcommand{\beQt}{\beta_{Q_2}}
\bigskip
\noindent
{\it A General Remark}

\medskip
From the above examples,
we extract the following general features of the Lax operator.
Let $M$ be a complex surface and $\tilde{M}$ be the blow up of $M$
at one point $p\in M$. We denote by $\pi:\tilde{M}\to M$ the projection,
and by $E$ the exceptional class $[\pi^{-1}(p)]$.
The second cohomology group of $\tilde{M}$ is given by \cite{GH}
\beq
H^2(\tilde{M})=\pi^*H^2(M)\oplus \Z E,
\eeq
which is an orthogonal decomposition with respect to the intersection form.
We choose a base $\{\omega_1,\ldots,\omega_r,E\}$
of $H^2(\tilde{M})$ 
so that $\omega_i\in \pi^*H^2(M)$ and
denote parameters of the marginal perturbations as
$t_1,\ldots,t_r,t_E$. We put $\beta_E=\e^{t_E}$ and
consider the limit $\beta_E\to 0$ with other parameters $t_i$ being fixed.
In the above examples of $M_2\to M_1$ and $M_1\to \CP^2$,
Lax operator $L_{\tilde{M}}$ and the fields of $\tilde{M}$
behave as
\beqa
L_{\tilde{M}}\,\,&\!\!\longrightarrow & \!L_M\\
&&\hspace{-1.5cm}\widehat{\pi^*O}\longrightarrow\widehat{O}\\
&&\hspace{-1.5cm}\widehat{E}/\beta_E\longrightarrow \widehat{R},
\eeqa
where $L_M$ and $O$ are the Lax operator and a primary field of $M$.
As a consequence, we have
\beqa
\lim_{\beta_E\to 0}\bra\sigma_n(R)\pi^*(O)\ket_{\tilde{M}}
&=&\bra\sigma_n(R)O\ket_M,
\label{bdf1}\\
\lim_{\beta_E\to 0}\frac{1}{\beta_E}\bra\sigma_n(R)E\ket_{\tilde{M}}
&=&\bra\sigma_n(R)R\ket_M.
\label{bdf2}
\eeqa
We conjecture that these features are generally true.
As a non-trivial check,
let us consider the case of $\pi:M_2\to\CP^1\times\CP^1$ (See Figure 1).
\vsp
        \begin{figure}[htb]
        \begin{center}
            \epsfxsize=4.5in\leavevmode\epsfbox{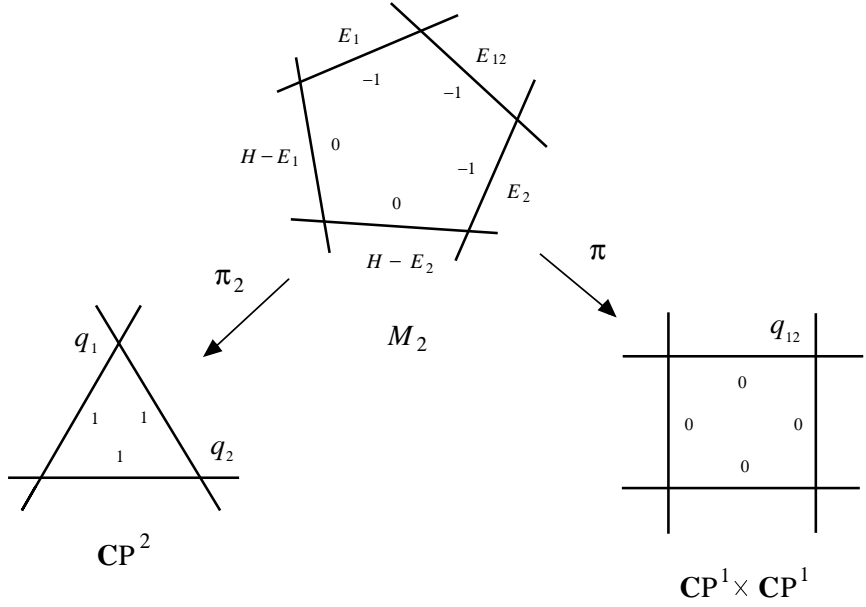}
        \end{center}
        \caption{Blow up of $\CP^2$ at two points $q_1$, $q_2$,
 and blow down along $E_{12}$}
        \end{figure}

The blow up $M_2$ of $\CP^2$ at two points $q_1$ and $q_2$
has three exceptional curves; $E_1=\pi_2^{-1}(q_1)$,
$E_2=\pi_2^{-1}(q_2)$, and $E_{12}=H-E_1-E_2$ as we noted before.
If we blow down along $E_{12}$,
we obtain the quadric surface $\CP^1\times\CP^1$.
The subspace of $H^2(M_2)$ orthogonal to $E_{12}$ is spanned by
$H-E_1=\tilde{Q}_2$ and $H-E_2=\tilde{Q}_1$.
Each of them has vanishing self-intersection and can be considered as 
the pull back of $Q_1$ or $Q_2$. 
Hence, it is appropriate to take the parametrization
\beqa
t_{Q_1}&=&t_H+t_{E_1}\\
t_{Q_2}&=&t_H+t_{E_2}\\
t_{E_{12}}&=&-t_H-t_{E_1}-t_{E_2},
\eeqa
and consider the limit $t_{E_{12}}\to -\infty$, or 
$\beta_{12}=\e^{t_{E_{12}}}\to 0$ with $t_{Q_1}$ and $t_{Q_2}$ being fixed.
We work in the asymmetric parametrization (\ref{Lax2pt2})
and make the change of variables
\beqa
X&=&\e^{t_H}p_1^{-1}p_2^{-1}\\
\tilY&=&\e^{t_H}p_1^{-1}.
\eeqa
Then, the Lax operator and the fields in $\pi^*H^*(\CP^1\!\times\!\CP^1)$
are expressed as
\beqa
L&=&\beQo \beQt\beta_{12}p_1^{-1}p_2^{-1}
+p_2+p_1+\beQo p_1^{-1}+\beQt p_2^{-1},\\[0.2cm]
&& \widehat{\tilde{Q}_1}=\beQo\beQt\beta_{12}p_1^{-1}p_2^{-1}
+\beQo p_1^{-1},\\
&& \widehat{\tilde{Q}_2}=\beQo\beQt\beta_{12}p_1^{-1}p_2^{-1}
+\beQt p_2^{-1},\\
&& \widehat{R}=\beQo\beQt\beta_{12}p_1^{-1}+\beQo p_1^{-1}p_2,
\nonumber
\eeqa
which have the well-defined limits as $\beta_{12}\to 0$:
\beqa
L&\to&p_2+p_1+\beQo p_1^{-1}+\beQt p_2^{-1}
\\[0.2cm]
&& \widehat{\tilde{Q}_1}\to \beQo p_1^{-1}\equiv p_1,\\
&& \widehat{\tilde{Q}_2}\to \beQt p_2^{-1}\equiv p_2 ,\\
&& \widehat{R}\to \beQo p_1^{-1}p_2\equiv p_1p_2 .
\eeqa
Thus, as $\beta_{12}\to 0$
the Lax operator and the fields become
those of the
$\CP^1\times\CP^1$ model (\ref{Laxquad}),(\ref{fieldsquad}) .
Note also that
$\widehat{E}_{12}=X=\beQo \beQt\beta_{12}p_1^{-1}p_2^{-1}$
and hence
\beq
\widehat{E}_{12}/\beta_{12}\to\beQo \beQt p_1^{-1}p_2^{-1}
\equiv p_1p_2.
\eeq
This shows that the conjecture holds in this case.

\bigskip

We have left many important probelms untouched in this paper. \\
\noindent (1): Our treatment of the Lax formulation is limited to the
case of marginal perturbations. Is it possible to generalize the
formalism to a larger phase space? \\
\noindent (2): We have restricted ourselves to the genus=0
case throughout
this paper (except the results of the $\CP^1$ model).
It is a challenging problem 
to generalize our discussions to higher genera. The known topological 
recursion relation for $g=1$ \cite{W1} unfortunately seems not
strong enough to determine genus=1 amplitudes except in the $\CP^1$ case.
Beyond $g=1$ no general relations are known for topological string
amplitudes. Is it possible that a suitable central 
extension of the Virasoro algebra which we encoutered in section 3
may describe the higher genus extention? \\
\noindent (3): We need a deeper understanding of the mirror phenomenon in
Fano varieties. \\
\noindent We hope to discuss these issues in a future publication.

\vspace{10mm}

 T.E. would like to thank P. Di Francesco for discussions at an early
stage of this work.
We also wish to thank R. Dijkgraaf, K. Intriligator, M. Jinzenji,
S.-K. Yang and especially I. Nakamura
for conversations, discussions and educations.
This research has been performed while C.S.X. was
a JSPS fellow at Univ. of Tokyo. Research of T.E. is suported by 
Grant-in-Aid for Scientific Research on Priority Area 213 
"Infinite Analysis", Japan Ministry of Education.

\newpage
\vspace{0.7cm}
\renewcommand{\theequation}{A.\arabic{equation}}\setcounter{equation}{0}

\noindent
{\large \bf Appendix A}

\medskip
In this appendix, we give a proof of the formula (\ref{LGtrr})
for the case of two variables.
We may put $t=t_0^P=0$.
First, we review the case of the $\CP^1$ model with
$L=p+p^{-1}$ \cite{EHY}.
Since we have
\beq
\frac{f(L)}{p}=\left(\frac{f(L)}{p}\right)_+
+\Bigl(f(L)\Bigr)_{\!0}\, p^{-1}+\Bigl(f(L)p\Bigr)_{\!0}\,p^{-2}+O(p^{-3}),
\eeq
and $p^2\partial_p L=p^2-1$ has only $+$ components,
we obtain
\beqa
\Bigl( f(L)p\partial_pL\Bigr)_+&=&\left(\frac{f(L)}{p}(p^2-1)\right)_+\\
&=&
\left(\frac{f(L)}{p}\right)_+(p^2-1)
+\Bigl(f(L)\Bigr)_{\!0}\, p+\Bigl(f(L)p\Bigr)_{\!0}\\
&\equiv&\Bigl(f(L)\Bigr)_{\!0}\, p+\Bigl(f(L)p\Bigr)_{\!0}.
\eeqa
The above argument applies also to multi-variable cases,
though we need more efforts. We present the derivation
in the case of $\CP^2$
of two variables.
Let us rename $Z_1$ as $Z$ and $Z_2$ as $W$.
The Lax operator is then rewritten as
$L=Z+W+Z^{-1}W^{-1}$ and we note that
\beq
Z^2W^2\,Z\partial_ZLW\partial_WL=Z^3 W^3-Z W^2-Z^2 W+1
\eeq
has only $+$ components.
The negative part of $f(L)/Z^2W^2$ can be expressed as
\beq
\left(\frac{f(L)}{Z^2 W^2}\right)_-:=
\frac{f(L)}{Z^2 W^2}-\left(\frac{f(L)}{Z^2 W^2}\right)_+
=
\left(
\begin{array}{c}
\\[-0.1cm]
{\displaystyle \sum}\\[-0.1cm]
{\scriptstyle a<0}\\[-0.2cm]
{\scriptstyle b\geq 0}
\end{array}
+
\begin{array}{c}
\\[-0.1cm]
{\displaystyle \sum}\\[-0.1cm]
{\scriptstyle a\geq0}\\[-0.2cm]
{\scriptstyle b< 0}
\end{array}
+
\begin{array}{c}
\\[-0.1cm]
{\displaystyle \sum}\\[-0.1cm]
{\scriptstyle a<0}\\[-0.2cm]
{\scriptstyle b<0}
\end{array}
\right)
\left(\frac{f(L)}{Z^{a+2}W^{b+2}}\right)_0 Z^a W^b.
\eeq
We note that $Z\equiv W$ and
$Z^3\equiv Z^2 W\equiv Z W^2\equiv W^3\equiv 1$
but we cannot use them before taking the $+$ part $(\cdots)_+$.
Taking this into account, we have
\beqa
\left(\left(\frac{f(L)}{Z^2 W^2}\right)_-
Z^3 W^3\right)_+&=&
\left(
\begin{array}{c}
\\[-0.1cm]
{\displaystyle \sum}\\[-0.1cm]
{\scriptstyle -a=1,2,3}\\[-0.2cm]
{\scriptstyle b\geq 0}
\end{array}
+
\begin{array}{c}
\\[-0.1cm]
{\displaystyle \sum}\\[-0.1cm]
{\scriptstyle a\geq0}\\[-0.2cm]
{\scriptstyle -b=1,2,3}
\end{array}
+
\begin{array}{c}
\\[-0.1cm]
{\displaystyle \sum}\\[-0.1cm]
{\scriptstyle -a=1,2,3}\\[-0.2cm]
{\scriptstyle -b=1,2,3}
\end{array}
\right)
\left(\frac{f(L)}{Z^{a+2}W^{b+2}}\right)_0Z^{a+3} W^{b+3}
\nonumber\\
&\equiv&\left(
\begin{array}{c}
\\[-0.1cm]
{\displaystyle \sum}\\[-0.1cm]
{\scriptstyle -a=1,2,3}\\[-0.2cm]
{\scriptstyle b\geq 0}
\end{array}
+
\begin{array}{c}
\\[-0.1cm]
{\displaystyle \sum}\\[-0.1cm]
{\scriptstyle a\geq0}\\[-0.2cm]
{\scriptstyle -b=1,2,3}
\end{array}
+
\begin{array}{c}
\\[-0.1cm]
{\displaystyle \sum}\\[-0.1cm]
{\scriptstyle -a=1,2,3}\\[-0.2cm]
{\scriptstyle -b=1,2,3}
\end{array}
\right)
\left(\frac{f(L)}{Z^{a+2}W^{b+2}}\right)_0Z^a W^b,
\nonumber\\
\left(\left(\frac{f(L)}{Z^2 W^2}\right)_-
Z W^2\right)_+&\equiv&
\left(
\begin{array}{c}
\\[-0.1cm]
{\displaystyle \sum}\\[-0.1cm]
{\scriptstyle a=-1}\\[-0.2cm]
{\scriptstyle b\geq 0}
\end{array}
+
\begin{array}{c}
\\[-0.1cm]
{\displaystyle \sum}\\[-0.1cm]
{\scriptstyle a\geq0}\\[-0.2cm]
{\scriptstyle b=-1,-2}
\end{array}
+
\begin{array}{c}
\\[-0.1cm]
{\displaystyle \sum}\\[-0.1cm]
{\scriptstyle a=-1}\\[-0.2cm]
{\scriptstyle b=-1,-2}
\end{array}
\right)
\left(\frac{f(L)}{Z^{a+2}W^{b+2}}\right)_0Z^a W^b
\nonumber\\
&\equiv&
\left(\left(\frac{f(L)}{Z^2 W^2}\right)_-
Z^2 W\right)_+\nonumber\\
\left(\left(\frac{f(L)}{Z^2 W^2}\right)_-\cdot 1\right)_+
&\equiv&0.\nonumber
\eeqa
Then, we have
\beqa
\lefteqn{
\left(\left(\frac{f(L)}{Z^2 W^2}\right)_-
Z^2W^2\,Z\partial_ZLW\partial_WL
\right)_+}
\label{naganaga}\\
&&\equiv
\left(
2
\begin{array}{c}
\\[-0.1cm]
{\displaystyle \sum}\\[-0.1cm]
{\scriptstyle a=-3}\\[-0.2cm]
{\scriptstyle b\geq 0}
\end{array}
+
\begin{array}{c}
\\[-0.1cm]
{\displaystyle \sum}\\[-0.1cm]
{\scriptstyle a=-2,-3}\\[-0.2cm]
{\scriptstyle b=-2,-3}
\end{array}
+
2
\begin{array}{c}
\\[-0.1cm]
{\displaystyle \sum}\\[-0.1cm]
{\scriptstyle a=-3}\\[-0.2cm]
{\scriptstyle b=-1}
\end{array}
-
\begin{array}{c}
\\[-0.1cm]
{\displaystyle \sum}\\[-0.1cm]
{\scriptstyle a=-1}\\[-0.2cm]
{\scriptstyle b=-1}
\end{array}
-
2
\begin{array}{c}
\\[-0.1cm]
{\displaystyle \sum}\\[-0.1cm]
{\scriptstyle a=-1}\\[-0.2cm]
{\scriptstyle b\geq 0}
\end{array}
\right)
\left(\frac{f(L)}{Z^{a+2}W^{b+2}}\right)_0Z^a W^b.
\nonumber
\eeqa
Here, we note that $L$ is invariant under the exchange of $Z$ and $W$,
and also under $Z\to Z^{-1}W^{-1}$, $W\to W$. Hence,
\beq
\Bigl(f(L)Z^{-a-2}W^{-b-2}\Bigr)_0=
\Bigl(f(L)Z^{a+2}W^{a-b}\Bigr)_0.
\eeq
In particular, 
$\Bigl(f(L)/Z^{a+2}W^{b+2}\Bigr)_0Z^aW^b$
is invariant (at the critical points) under
\beq
(a,b)=(-3,b)\to (-1,b+1).
\eeq
Then, it is easy to see that (\ref{naganaga})
acquires contribution from
$(a,b)=(-2,-2)$, $(-2,-3)$ and $(-3,-3)$, all with multiplicity 1.
Namely,
\beqa
\lefteqn{
\left(\left(\frac{f(L)}{Z^2 W^2}\right)_-
Z^2W^2\,Z\partial_ZLW\partial_WL
\right)_+
}\\
&&\equiv
\Bigl(f(L)\Bigr)_0Z^{-2}W^{-2}+
\Bigl(f(L)W\Bigr)_0Z^{-2}W^{-3}+
\Bigl(f(L)ZW\Bigr)_0Z^{-3}W^{-3}
\\
&&\equiv
\Bigl(f(L)\Bigr)_0ZW+
\Bigl(f(L)Z\Bigr)_0W+
\Bigl(f(L)ZW\Bigr)_0.
\eeqa
Since
$\Bigl(
\Bigl(f(L)/Z^2W^2\Bigr)_+Z^2W^2\,Z\partial_ZLW\partial_WL\Bigr)_+
\equiv 0$,
we see that (\ref{LGtrr}) holds for $N=2$.

\newpage
\vspace{0.7cm}
\newcommand{\uu}{_{{\scriptstyle \Box}\hspace{-0.069cm}{\scriptstyle \Box}}}
\newcommand{\due}{\hspace{-0.17cm}\begin{array}{c}
\\[-0.27cm]
{\scriptstyle \Box}\\[-0.368cm]
{\scriptstyle \Box}
\end{array}
\!\!}
\newcommand{\du}{\hspace{-0.17cm}\begin{array}{cc}
&\\[-0.27cm]
{\scriptstyle \Box}&\hspace{-0.63cm}{\scriptstyle \Box}\\[-0.368cm]
{\scriptstyle \Box}&
\end{array}
\!\!\!\!\!}
\newcommand{\dd}{\hspace{-0.17cm}\begin{array}{cc}
&\\[-0.27cm]
{\scriptstyle \Box}&\hspace{-0.63cm}{\scriptstyle \Box}\\[-0.368cm]
{\scriptstyle \Box}&\hspace{-0.63cm}{\scriptstyle \Box}
\end{array}
\!\!\!\!\!}
\newcommand{\tre}{\hspace{-0.17cm}\begin{array}{c}
\\[-0.27cm]
{\scriptstyle \Box}\\[-0.365cm]
{\scriptstyle \Box}\\[-0.366cm]
{\scriptstyle \Box}
\end{array}
\!\!}
\newcommand{\tu}{\hspace{-0.17cm}\begin{array}{cc}
&\\[-0.27cm]
{\scriptstyle \Box}&\hspace{-0.63cm}{\scriptstyle \Box}\\[-0.365cm]
{\scriptstyle \Box}&\\[-0.366cm]
{\scriptstyle \Box}&
\end{array}
\!\!\!\!\!}
\newcommand{\td}{\hspace{-0.17cm}\begin{array}{cc}
&\\[-0.27cm]
{\scriptstyle \Box}&\hspace{-0.63cm}{\scriptstyle \Box}\\[-0.365cm]
{\scriptstyle \Box}&\hspace{-0.63cm}{\scriptstyle \Box}\\[-0.366cm]
{\scriptstyle \Box}&
\end{array}
\!\!\!\!\!}
\newcommand{\tretre}{\hspace{-0.17cm}\begin{array}{cc}
&\\[-0.27cm]
{\scriptstyle \Box}&\hspace{-0.63cm}{\scriptstyle \Box}\\[-0.365cm]
{\scriptstyle \Box}&\hspace{-0.63cm}{\scriptstyle \Box}\\[-0.366cm]
{\scriptstyle \Box}&\hspace{-0.63cm}{\scriptstyle \Box}
\end{array}
\!\!\!\!\!}

\medskip
\bigskip
\renewcommand{\theequation}{B.\arabic{equation}}\setcounter{equation}{0}

\noindent
{\large \bf Appendix B:\hspace{0.3cm}Grassmannians}

\bigskip
Cohomology group of the complex Grassmann manifold $Gr(k,N)$ (the space of
$k$-dimensional subspaces in $\C^N$) is spanned by
Schubert classes which are in one to one correspondence
with Young diagrams of height $\leq N-k$ and width $\leq k$.
We denote by $O_{\rm Y}$ the primary field corresponding
to a Young diagram ${\rm Y}$.
Pairing of $O_{\rm Y}$ and $O_{\rm Y^{\prime}}$
is nonvanishing (and is $1$)
if and only if the union of ${\rm Y}$ and ${\rm Y^{\prime}}$ form
the rectangular diagram of size $(N-k)\times k$.
The dimension is $k(N-k)$ and
the first Chern class is given by $N O_{\Box}$.

\bigskip
\noindent
{\sc B.1 Lax Operators}

\bigskip
\noindent
\underline{$Gr(2,4)$}

\medskip
First, we consider the simplest case $Gr(2,4)$.
There are six primaries
$P$, $O_{\Box}$, $O\due$, $O\uu$, $O\du$, $O\dd$.
The fundamental matrix in the marginal perturbation $t_0^{\Box}=t$
is given by
\beq
\left(
\begin{array}{cccccc}
0&4&0&0&0&0\\
0&0&4&4&0&0\\
0&0&0&0&4&0\\
0&0&0&0&4&0\\
4\e^t\!\!&0&\,0\,&\,0\,&\,0\,&\,4\,\\
0&4\e^t&0&0&0&0
\end{array}
\right).
\eeq
Here we have used the result for the
instanton contribution $\bra O_{\Box}O\du O\dd\ket=\e^t$
which will be derived in the following for
a more general situation. (See also \cite{IV,W4,Bert,Bert2}.)
We find that the Lax operator and the fields are given by
\beqa
L&=&X+X^{-1}(Y+Z)+(Y^{-1}+Z^{-1})W+\e^tW^{-1},\\[0.2cm]
&&\widehat{P}=1\label{Bid1}\\
&&\widehat{O}_{\Box}=X\\
&&\widehat{O}\due=Y,\quad\widehat{O}\uu=Z\\[-0.2cm]
&&\widehat{O}\du=(1+YZ^{-1})W\\[-0.2cm]
&&\widehat{O}\dd=X W\label{Bid5}
\eeqa

\bigskip
\noindent
\underline{$Gr(2,5)$}

\medskip
There are ten primaries;
$P$, $O_{\Box}$, $O\due$, $O\uu$, $O\tre$, $O\du$, $O\tu$,
$O\dd$, $O\td$, $O\tretre$.
The fundamental matrix in the marginal perturbation is given
by
\beq
\left(
\begin{array}{cccccccccc}
0&5&0&0&0&0&0&0&0&0\\
0&0&5&5&0&0&0&0&0&0\\
0&0&0&0&5&5&0&0&0&0\\
0&0&0&0&0&5&0&0&0&0\\
0&0&0&0&0&0&5&0&0&0\\
0&0&0&0&0&0&5&5&0&0\\
\!5\e^t\!\!&0&0&0&0&0&0&0&5&0\\
0&0&0&0&0&0&0&0&5&0\\
0&\!\!5\e^t\!\!&0&0&0&0&0&0&0&5\\
0&0&\!\!5\e^t\!\!&0&0&0&0&0&0&0
\end{array}
\right).
\eeq
The Lax operator and the fields are given by
\beqa
L&=&X+X^{-1}(Y+Z)+Y^{-1}W+(Y^{-1}+Z^{-1})U+(W^{-1}+U^{-1})V+\e^tV^{-1},
\\[0.2cm]
&&\widehat{P}=1\\
&&\widehat{O}_{\Box}=X\\
&&\widehat{O}\due=Y,\quad\widehat{O}\uu=Z\\[-0.2cm]
&&\widehat{O}\tre=W,\quad\widehat{O}\du=(1+YZ^{-1})U\\[-0.2cm]
&&\widehat{O}\dd=XU,\quad\widehat{O}\tu=(1+WU^{-1})V\\[-0.2cm]
&&\widehat{O}\td=(1+UW^{-1})XV\\[-0.2cm]
&&\widehat{O}\tretre=YV.
\eeqa

\bigskip
\noindent
\underline{General $Gr(k,N)$}

\medskip

In the above two examples, we observe
a certain relation between the fundamental matrix and the Lax operator.
We shall call a Young diagram {\it slim} when
it has at most one row and one column.
We assign a variable $X_{\rm Y}$ to each slim diagram ${\rm Y}$,
and we put $X_{\rm empty}=1$. Then, in the above two examples
we find
\beq
L\,\,=\,\,\frac{1}{N}\sum_{{\rm slim\,\, diagrams}}
X_{\rm Y}^{-1}M_{\rm Y}^{\,\,\rm Y^{\prime}}X_{\rm Y^{\prime}}.
\label{Lgra}
\eeq
We conjecture that this generally holds.
In order to express it more explicitly,
we denote by $[a,b]$ the slim diagram of height $a$ and width $b$:
\beq
[a,b]\,=\,
a\left\{
\begin{array}{c}
\\
\\[0.15cm]
\end{array}\right.\!\!\!\!\!
\overbrace{
\begin{array}{ccc}
\!\!\!\!\Box&\hspace{-0.59cm}\Box&\hspace{-1cm}\cdot\!\cdot\!\cdot
\Box\!\!\!\!\!\!\!\!\!\!\\[-0.297cm]
\!\!\!\!\Box&&\\[-0.297cm]
\!\!\!\!\vdots&&\\[-0.15cm]
\!\!\!\!\Box&&
\end{array}}^{\displaystyle b}
\eeq
First, we determine the minor $M_{\rm Y}^{\,\,\rm Y^{\prime}}$.
By the selection rule,
\beq
\bra O_{\rm Y} O^{\rm Y^{\prime}}\ket_d\ne 0\quad\Longrightarrow
\quad Nd+k(N-k)-1=|{\rm Y}|+k(N-k)-|{\rm Y^{\prime}}|,
\eeq
where $|{\rm Y}|$ is the number of boxes of ${\rm Y}$.
Since $|{\rm Y}|\leq N-1$ for slim diagrams, we see that the degree
must be $0$ or $1$ and $d=1$ occurs only for ${\rm Y}=[N-k,k]$,
and ${\rm Y^{\prime}}=[0,0]$.
Let us consider the latter case of degree-1 instanton contribution.
The Poincar\'e dual of $O^{\rm Y^{\prime}}$ is a point $p$ and that
of $O_{\rm Y}$ is represented by the closure of a Schubert cell
$C_{\rm Y}\subset Gr(k,N)$. We choose as the point $p$ the subspace
spanned by $e_{N-k+1},\ldots,e_{N}$,
where $e_1,\ldots,e_N$ is the standard base of $\C^N$.
We choose as the cell $C_{\rm Y}$ the $B_+$-orbit of the point
corresponding to the subspace spanned by
$e_1,e_{N-k+1},\ldots,e_{N-1}$, where $B_+$ is the subgroup of
$GL(N,\C)$ consisting of upper triangular matrices.
A degree one map $\CP^1\to Gr(k,N)$ is given by a family
$(s:t)\mapsto W(s:t)$ of subspaces of the form
\beq
W(s:t)=\{v_0s+v_1t,v_2,\ldots,v_k\}_{\C},
\eeq
where $v_0,\ldots,v_k$ are linearly independent elements of $\C^N$.
Here we consider $(s,t)$ as the homogeneous coordinates of $\CP^1$.
Let $(0:1)$ and $(1:0)$ be the insertion point of $O^{\rm Y^{\prime}}$
and $O_{\rm Y}$ respectively. We count the number of families
(up to the automorphism $\C^*$ of $\CP^1-\{(0:1),(1:0)\}$)
such that $W(0:1)=p$ and $W(1:0)\in\overline{C_{\rm Y}}$.
The requirements are restated as
\beqa
\{v_1,v_2,\ldots,v_k\}_{\C}&=&\{e_{N-k+1},\ldots,e_{N}\}_{\C},\hspace{2cm}\\
e_1\in\{v_0,v_2,\ldots,v_k\}_{\C}&\subset&\{e_1,\ldots,e_{N-1}\}_{\C}.
\eeqa
Then, we find
\beq
W(s:t)=\{e_1cs+e_Nt,e_{N-k+1},\ldots,e_{N-1}\}_{\C},
\eeq
where $c\in\C^*$. Thus, the number is one up to the $\C^*$ action:
\beq
\bra O_{\rm Y} O^{\rm Y^{\prime}}\ket_1=1, \hskip3mm Y=[N-k,k],\hskip1mm
Y'=[0,0]
\eeq

The classical part of the cohomology ring is well-known.
Finally, we can express the conjectured form (\ref{Lgra})
for the Lax operator:
\beq
L\,=\,X_{[1,1]}\,+\sum_{{\scriptstyle 1\leq a\leq N-k}
\atop {\scriptstyle 1\leq b\leq k}}X_{[a,b]}^{-1}(X_{[a+1,b]}+X_{[a,b+1]})
\,+\,\e^t X_{[N-k,k]}^{-1},
\eeq
where we put $X_{[a,b]}=0$ if $a>N-k$ or $b>k$.
Note that the number of variables or the number
of non-empty slim diagrams is $(N-k)\times k$,
which is the same as the dimension of the original target space $Gr(k,N)$.

\bigskip
\noindent
{\sc B.2 Landau-Ginzburg Description}

\medskip
Finally, let us discuss the LG description in the simplest case $Gr(2,4)$.
The equation determining the vacua
$X\partial_XL=Y\partial_YL=Z\partial_ZL=W\partial_WL=0$
has four solutions
\beq
Y=Z=\frac{1}{2}X^2,\quad
W=\frac{1}{4}X^3,\quad X^4=4\e^t,
\eeq
each having non-vanishing Hessian $8\e^t$.
Four is less than the index $\chi(Gr(2,4))=6$,
and there are vacua at infinity:
$X\to 0$ keeping $XW=Y^2$ and $Z+Y=0$.
Namely, the LG model based on the algebraic torus
$(\C^*)^4=\{(X,Y,Z,W)\}$ with superpotential $L$ and holomorphic form
$\Omega=\frac{dX}{X}\wedge\frac{dY}{Y}\wedge\frac{dZ}{Z}\wedge\frac{dW}{W}$
has a disease and can not be considered as a sound description
of the original system.
To obtain a good one,
we must partially compactify the torus
so that $L$ and $\Omega$ are extended and there are six vacua in total.
We make the change of variables $X,Y,Z,W\to X,Y,\zeta,\tilde{W}$
in a neighborhood of the ``vacua at infinity'' $X=Y+Z=W^{-1}=0$:
\beq
W=X^{-1}\tilde{W},\qquad Z+Y=-XY\zeta,
\eeq
and include the points with $X=0$.
Then, we obtain a new expression
\beqa
L&=&X-Y\zeta+Y^{-1}(1+X\zeta)^{-1}\zeta \tilde{W}
+\e^tX\tilde{W}^{-1},\\[0.2cm]
\Omega&=&dX\wedge\frac{dY}{Y}
\wedge \frac{d\zeta}{1+X\zeta}\wedge\frac{d\tilde{W}}{\tilde{W}}.
\eeqa
Solving the equation
$\partial_XL=Y\partial_YL
=\partial_{\zeta}L=\tilde{W}\partial_{\tilde{W}}L=0$,
we find two additional vacua
\beq
X=\zeta=0,\qquad Y^2=\tilde{W}=-\e^t,
\eeq
each having non-vanishing Hessian $-4\e^t$.
In total there are $4+2=6$ vacua and there are no more at infinity.
Moreover, it is easy to see that the three point functions
of the original model
coincide with those of this LG model
under the identification of fields (\ref{Bid1})-(\ref{Bid5}).
In summary, we have found a sound LG description of the $Gr(2,4)$ model
based on a partial compactification of $(\C^*)^4$.
This is the first example in which the mirror partner
of a Fano manifold is not just an algebraic torus.

\end{document}